\documentclass[12pt,a4paper]{JHEP3}
\usepackage[all]{xy}
\usepackage{graphics}
\usepackage{epsfig}

\def\p{\partial}

\def\a{\alpha}
\def\b{\beta}
\def\g{\gamma}
\def\d{\delta}
\def\e{\epsilon}
\def\k{\kappa}
\def\m{\mu}

\def\l{\lambda}

\def\s{\sigma}
\def\p{\partial}
\def\t{{\theta}}
\def\T{{\Theta}}
\def\S{\Sigma}

\def\O{\Omega}
\def\G{\Gamma}

\def\z{\zeta}

\newcommand{\Tr}{{\textrm{Tr}}}

\newcommand{\be}{\begin{equation}}
\newcommand{\ee}{\end{equation}}
\newcommand{\bea}{\begin{eqnarray}}
\newcommand{\eea}{\end{eqnarray}}
\newcommand{\Str}{\ensuremath{\mathrm{Str}}}

\title{Self-Duality of Green-Schwarz Sigma-Models}

\author{Amit~Dekel and Yaron~Oz \\
  Raymond and Beverly Sackler School of Physics and Astronomy \\
  Tel-Aviv University, Ramat-Aviv 69978, Israel\\
\email{amitde@post.tau.ac.il}, \email{yaronoz@post.tau.ac.il}}

\abstract{
  We study fermionic T-duality symmetries of integrable Green-Schwarz sigma-models on Anti-de-Sitter backgrounds with Ramond-Ramond
  fluxes, constructed as $\mathbb{Z}_4$ supercosets of superconformal algebras.
  We find three algebraic conditions that guarantee self-duality of the backgrounds under fermionic T-duality, we classify those that satisfy them
  and construct the map of the monodromy matrix.
  We introduce new T-duality directions, where some of them contain no bosonic directions,
  along which the backgrounds are self-dual.
  We find that the only self-dual backgrounds are $\mathrm{AdS}_n\times \mathrm{S}^n$ for $n=2,3,5$. In addition we find that the backgrounds
   $\mathrm{AdS}_n\times \mathrm{S}^1$ for $n=2,3,5$, $\mathrm{AdS}_4\times \mathrm{S}^2$ and $\mathrm{AdS}_2\times \mathrm{S}^4$
  are self-dual at the level of the classical action, but have a non-trivial transformation of the dilaton.}


\preprint{}

\begin{document}
\section{Introduction}
Self-duality of the Green-Schwarz sigma model (GSSM) on $\mathrm{AdS}_5\times \mathrm{S}^5$ background is used to explain the existence of the dual-superconformal symmetry
of scattering amplitudes in $\mathcal{N}=4$ SYM, and their connection to Wilson-loops \cite{Berkovits:2008ic}\cite{Beisert:2008iq}.
The superconformal symmetry together with the dual one generate  a Yangian symmetry algebra, which is related to the integrability properties of the theory.

It is well known that GSSM's on semi-symmetric spaces ($\mathbb{Z}_4$ supercoset spaces) exhibit an infinite set of conserved charges \cite{Bena:2003wd} which satisfy the Yangian algebra \cite{Dolan:2003uh}\cite{Dolan:2004ps}.
It is thus natural to ask whether GSSM's on other (than $\mathrm{AdS}_5\times \mathrm{S}^5$) semi-symmetric backgrounds are self-dual under T-duality.
In previous papers \cite{Adam:2009kt}\cite{Adam:2010hh}\cite{Hao:2009hw}, some backgrounds were checked to be self-dual, while other were found not to be self-dual.
In those papers, the background's self-duality was checked on a case by case basis. A general argument for self-duality is still lacking.
In the present paper we will take a rather general approach and formulate criteria for semi-symmetric backgrounds to be self-dual.
We present three sufficient algebraic conditions for self-duality, and explain the lack of self-duality of backgrounds that do not satisfy them.

We denote the superconformal algebras (SCA's) by $\mathfrak{g}$, with the $\mathbb{Z}_2$ decomposition $\mathfrak{g}=\mathfrak{g}_{\bar 0}\oplus \mathfrak{g}_{\bar 1}$ to its even and parts respectively.
We further decompose the SCA's according to a $\mathbb{Z}$-gradation with gradings $\pm 1,0$ only, where the charges are assigned by a generator $U$. The T-duality is performed along all the directions associated with the grading $1$ generators, which form an abelian subalgebra. We will prove that a background is self-dual
if :
\begin{enumerate}
  \item $\O(U)=-U$, where $\O$ is the $\mathbb{Z}_4$ automorphism map.
  \item Rank($\k$-symmetry) $\geq$ dim$(\mathfrak{g}_{\bar 1})$/4.
  \item The SCA's Killing-form vanishes.
\end{enumerate}

The first condition ensures a non-singular coupling of the fermionic coordinates.
The second condition allows a particular representation of the supergroup that is used in the T-duality procedure. The third condition guarantees
the quantum consistency of the transformation, that is a non-trivial  dilaton is not generated.

We find that the only self-dual GSSM's are the $\mathrm{AdS}_n\times \mathrm{S}^n$ for $n=2,3,5$.  All of them were found previously to be self-dual \cite{Berkovits:2008ic}\cite{Beisert:2008iq}\cite{Adam:2009kt}. We find there are also backgrounds that are self-dual at the classical level, but at the quantum level their dilaton shifts, these are the $\mathrm{AdS}_n\times \mathrm{S}^1$ for $n=2,3,5$ (the case of $n=5$ was discussed in \cite{Hao:2009hw}), $\mathrm{AdS}_2\times \mathrm{S}^4$, and $\mathrm{AdS}_4\times \mathrm{S}^2$.
In addition to the usual self-duality along the flat AdS directions followed by some odd directions, namely the directions associated with span$\{P,Q\}$, we find other abelian subalgebras along which the GSSM is self-dual (one of them was discussed in \cite{Berkovits:2008ic}). Some of these directions involve only fermionic directions.
 We give the general transformation of the action and the flat-connection for any such abelian subalgebra. The transformation, as in the $\mathrm{AdS}_5\times \mathrm{S}^5$ case, is a spectral parameter dependent automorphism, which is a composition of the $\mathbb{Z}_4$-automorphism map and an automorphism induced by the $\mathbb{Z}$-gradation.

The paper is organized as follows. In section \ref{sec:Properties of Superconformal Algebras} we briefly review some properties of the SCA's, including a discussion of their $\mathbb{Z}$-gradation structure. In section \ref{sec:Green-Schwarz Sigma-models on Semi-Symmetric backgrounds} we briefly discuss the GSSM and their basic integrability properties. In section \ref{sec:T-duality of the GS Sigma-Models} we prove T-self-duality of the GSSM's using the three algebraic conditions stated above.
In section \ref{sec:Classification of the backgrounds} we classify the SCA's according to the conditions for T-self-duality. In section \ref{sec:disscasion} we discuss the results and various open questions. In appendix \ref{sec:Notations} we summarize our notations. In appendix \ref{ap:Superalgebras} we provide technical computations concerning the SCA's and their classification according to the first condition. In appendix \ref{ap:kappa} we compute the kappa-symmetry needed for the second condition.

\section{Properties of Superconformal Algebras}\label{sec:Properties of Superconformal Algebras}
\subsection{The conformal basis and $\mathbb{Z}$-gradation}
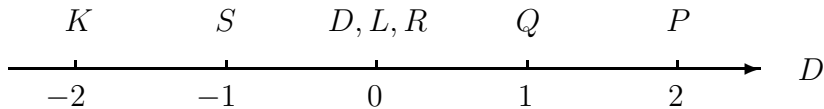
\begin{figure*}[!ht]
\centering
\setlength{\unitlength}{1cm}
\begin{picture}(10,2)
\thicklines
\put(0,0){$-2$}
\put(2,0){$-1$}
\put(4,0){$~~0$}
\put(6,0){$~~1$}
\put(8,0){$~~2$}
\put(0.25,1){$K$}
\put(2.25,1){$S$}
\put(3.75,1){$D,L,R$}
\put(6.25,1){$Q$}
\put(8.25,1){$P$}
\put(-0.5,0.5){\vector(1,0){10}}
\put(10,0.35){$D$}
\put(0.4,0.5){\line(0,1){0.1}}
\put(2.4,0.5){\line(0,1){0.1}}
\put(4.4,0.5){\line(0,1){0.1}}
\put(6.4,0.5){\line(0,1){0.1}}
\put(8.4,0.5){\line(0,1){0.1}}
\end{picture}
\caption{The charge of the SCA's generators under $D$.}
\label{fig:Dcharge}
\end{figure*}

The generators of the SCA in $d$-dimensions are $\mathfrak{g}_\mathrm{C}=\mathrm{span}\{P, K, D, L\}$ - the $\mathrm{so}(2,d-1)$ conformal subalgebra generators, $\mathrm{span}\{R\}$ - the R-symmetry subalgebra generators,
and $\mathrm{span}\{Q\}$ and $\mathrm{span}\{S\}$ - the (odd) supercharges and superconformal charges respectively.
Altogether we have $\mathfrak{g}_{\mathrm{SC}}=\mathrm{span}\{P,K,L,D;R;Q,S\}$.
The SCA's super-commutation relations are given by the commutation relations of $\mathfrak{g}_\mathrm{C}$ and $\mathrm{span}\{R\}$ together with
\be\label{eq:SCAcommutationrelations}
[P,Q]=0,\quad
[K,S]=0,\quad
[P,S]\sim Q,\quad
[K,Q]\sim S,\quad
\ee
$$
[R,Q]\sim Q,\quad
[R,S]\sim S,\quad
$$
$$
\{Q,Q\}\sim P,\quad
\{S,S\}\sim K,\quad
$$
$$
\{Q,S\}\sim D+L+R.
$$
These commutation relations can summarized using the charge of the generators under the dilatation generator $D$, see figure \ref{fig:Dcharge}.
This charge assignment is an example of $\mathbb{Z}$-gradation of the SCA, which is a decomposition of the algebra such that
$\mathfrak{g}=\bigoplus_{i\in \mathbb{Z}}\mathfrak{g}_i$ and $[\mathfrak{g}_i,\mathfrak{g}_j]\subset\mathfrak{g}_{i+j}$.
In this case $i=-2,-1,0,1,2$.
Besides this $\mathbb{Z}$-gradation, the SCA may have others.

The superalgebras are classified according to their type, I or II \cite{Frappat:1996pb}.
The terminology type I and type II refers to the representation of the even part of the superalgebra on the odd part. If the representation is irreducible the superalgebra is called type II and if it is a direct sum of two irreducible representations the superalgebra is called type I.
In the case of type I, the odd part decomposes according to another $\mathbb{Z}$-gradation, which is called the \textit{distinguished gradation} \cite{Frappat:1996pb}. This gradation is associated with the generator $B$ (which we call the hypercharge) which is in the algebra for $A(m,n\neq m)$ and $C(n+1)$ and not for $A(m,m)$. The generators are decomposed as
\be\label{eq:distinguishedI}
\mathfrak{g}_I=\mathfrak{g}_1\oplus\mathfrak{g}_0\oplus\mathfrak{g}_{-1}=\{Q,\hat S\}\oplus\{P,K,D,L;R\}\oplus\{\hat Q,S\}.
\ee
For type II SCA's we do not have such a gradation, but when the number of space-time supersymmetries is even, $\mathcal{N}\in 2\mathbb{N}_1$,
we do have another $\mathbb{Z}$-gradation associated with a generator of the R-symmetry subalgebra which we call $\check \l$. This decomposition, which further decomposes the R-symmetry generators to $R,\l,\hat R$, is given by
\be\label{eq:distinguishedII}
\mathfrak{g}_{II}=\mathfrak{g}_2\oplus\mathfrak{g}_1\oplus\mathfrak{g}_0\oplus\mathfrak{g}_{-1}\oplus\mathfrak{g}_{-2}=\{R\}\oplus\{Q,\hat S\}\oplus\{P,K,D,L;\l\}\oplus\{\hat Q,S\}\oplus\{\hat R\}.
\ee
In order to present the SCA's, one has to work with real-forms of the SCA, since we have to take complex combinations of the odd generators and the R-symmetry generators.
We summarize some relevant properties of the SCA's in table \ref{table:SCA's}.
Further decomposition of the commutation relations (\ref{eq:SCAcommutationrelations}) should be obvious from the gradations introduced above. \begin{table}[ht]
\caption{Some properties of SCA's}
\centering
\scriptsize
\begin{tabular}{l l l l l l l}
\hline\hline
$d$ & SCA &  R-symmetry & dim($\mathfrak{g}_{\bar 1}$) & $\mathcal{N}$ & type & Killing-form\\ [0.5ex]
\hline \hline
1 & $\mathrm{osp}(N|2)$ & $\mathrm{so}(N)$ & $2N$ & $N$ & I for $N=2$, else II & ND, except for $N=4$ \\ \hline
1 & $\mathrm{su}(1,1|N\neq 2)$ & $\mathrm{u}(N)$ & $4N$ & $2N$& I & ND \\ \hline
1 & $\mathrm{psu}(1,1|2)$ & $\mathrm{su}(2)$ & $8$ & $4$& I & Zero \\ \hline
1 & $\mathrm{osp}(4^*|2N)$ & $\mathrm{su}(2)\times \mathrm{usp}(2N)$ & $8N$ & $4N$& II & ND \\ \hline
1 & $\mathrm{G}(3)$ & $\mathrm{g}_2$ & $14$ & $7$& II & ND \\ \hline
1 & $\mathrm{F}(4;0)$ & $\mathrm{so}(7)$ & $16$ & $8$& II & ND \\ \hline
1 & $\mathrm{D}(2,1;\a)$ & $\mathrm{so}(4)$ & $8$ & $4$& II & Zero \\ \hline
3 & $\mathrm{osp}(N|4)$ & $\mathrm{so}(N)$ & $4N$ & $N$& I for $N=2$, else II & ND, except for $N=6$ \\ \hline
4 & $\mathrm{su}(2,2|N\neq 4)$ & $\mathrm{u}(N)$ & $8N$ & $N$& I & ND \\ \hline
4 & $\mathrm{psu}(2,2|4)$ & $\mathrm{su}(4)$ & $32$ & $4$& I & Zero \\ \hline
5 & $\mathrm{F}(4;2)$ & $\mathrm{su}(2)$ & $16$ & $2$& II & ND \\ \hline
6 & $\mathrm{osp}(8^*|N)$ & $\mathrm{usp}(N)$ & $8N$ & $N$& II & ND, except for $N=6$ \\ \hline
\end{tabular}
\begin{quote}
The table gives the SCA's as classified in \cite{Nahm:1977tg}. The spinor representations for $d=3,4,5,6$ are $\mathrm{su}(2),\mathrm{su}(2)\times \mathrm{su}(2),\mathrm{sp}(4),\mathrm{su}(4)$ respectively. ND- stands for non-degenerate. $\mathcal{N}$ is the number of space-time supersymmetries.
\end{quote}
\label{table:SCA's}
\end{table}
\normalsize

Another characteristic of the SCA's is whether the Killing-form is degenerate or not, see table \ref{table:SCA's}. The Killing-form is defined as the supertrace of every two generators in the adjoint representation \cite{Kac:1977em}\cite{Frappat:1996pb}, $K_{ab}=\Str(L_a^{\mathrm{adj}} L_b^{\mathrm{adj}})$. The SCA's with degenerate Killing-form are known to have special properties in the context of the Green-Schwarz sigma-models, e.g \cite{Berkovits:1999zq}\cite{Kagan:2005wt}\cite{Zarembo:2010sg}, and as we shall see they are also special with respect to the self-duality properties of the sigma-models.

\subsection{$\mathbb{Z}_4$ automorphism}\label{sec:Z4automorphism}
Every SCA has at least one $\mathbb{Z}_4$ automorphism \cite{Serganova83}. A SCA is decomposed under this automorphism into four sets
\be
\mathfrak{g}=\mathcal{H}_0\oplus\mathcal{H}_1\oplus\mathcal{H}_2\oplus\mathcal{H}_3,
\ee
such that $[\mathcal{H}_i,\mathcal{H}_j\}\subset \mathcal{H}_{i+j~\mathrm{mod}~4}$, $B(\mathcal{H}_i,\mathcal{H}_j)\neq 0$ only if $i+j=0~\mathrm{mod}~4$, and $\O(\mathcal{H}_k)=i^k\mathcal{H}_k$, where $B$ represents the Cartan-Killing bilinear-form and $\O(\cdot)$ is the automorphism map.

Using the $\mathbb{Z}_4$ automorphism property we can define a semi-symmetric space by taking the quotient with respect to the invariant locus $\mathcal{H}_0$ (so the bosonic part is a symmetric-space).
A SCA may have several different $\mathbb{Z}_4$ automorphisms and so one can identify a semi-symmetric space with respect to each automorphism. Some of the semi-symmetric spaces will have a bosonic AdS sub-space in which we are mainly interested in the present paper, although we will also consider non-AdS spaces.

\subsection{More on $\mathbb{Z}$-gradations}\label{sec:Z-gradation}

As we have shown above, any SCA has a $\mathbb{Z}$-gradation, $\mathfrak{g}=\bigoplus_{i\in \mathbb{Z}}\mathfrak{g}_i$ such that $[\mathfrak{g}_i,\mathfrak{g}_j]\subset\mathfrak{g}_{i+j}$.
When $i$ takes a finite number of values, say $i_{\mathrm{min}}\leq i\leq i_{\mathrm{max}}$,
the set $\bigoplus_{i=i_{\mathrm{max}}/2}^{i_{\mathrm{max}}}\mathfrak{g}_{i}$ defines an abelian subalgebra if $i_{\mathrm{max}}>0$, and similarly for the set $\bigoplus_{i=i_{\mathrm{min}}}^{i_{\mathrm{min}}/2}\mathfrak{g}_{i}$ if $i_{\mathrm{min}}<0$.

In the present paper we will be interested in $\mathbb{Z}$-gradations with $|i|\leq 1$,
that is $\mathfrak{g}=A_1\oplus B_0\oplus A_{-1}$, with $A_{\pm 1}$ abelian subalgebras\footnote{The $\mathbb{Z}$-gradation considered should not necessarily be consistent, namely $\mathfrak{g}_{\pm 1}$ and $\mathfrak{g}_0$ may contain even and odd generators respectively \cite{Kac:1977em}\cite{Frappat:1996pb}.}
and $B_0$ a subalgebra, so these are the $\mathbb{Z}$-gradations we will consider from now on.
Any such decomposition can be induced by introducing a U$(1)$ generator $U$, with respect to $\mathfrak{g}_0$ (which may or may not be part of the SCA), satisfying $\mathrm{ad}_U(L_a)=[U,L_a]=a L_a$, $\forall$ $L_a\in \mathfrak{g}$, where $a=\pm 1,0$ is the charge of the generator.
The decomposition under $\mathrm{ad}_U$ induces a one-parameter dependent automorphism
\be\label{eq:Zgradationautomorphism}
\s_\l(L_a)=\l^{U} L_a \l^{-U}= \l^a L_a,\quad \l\in \mathbb{C}.
\ee

The various $\mathbb{Z}_4$ automorphisms \cite{Serganova83} of the SCA's may have different relations with the U$(1)$ generator inducing the $\mathbb{Z}$-gradation.
We are interested in those  satisfying
\be\label{eq:Z4Z3condition}
\O(U)=-U.
\ee
$\O$ is defined to act on a commutator as $\O([L_a,L_b])=[\O(L_a),\O(L_b)]$, thus
\be
\s_\l(\O(L_a))=\l^{-a}(\O(L_a)),
\ee
and the non-trivial bilinear-form is of the form $B(L_a\O(L_b))$ with $a=b$.
For backgrounds satisfying (\ref{eq:Z4Z3condition}) and some other conditions (to be discussed later) we will be able to prove T-self-duality.

Next we consider four classes of $\mathbb{Z}$-gradations that will be of interest in the study of T-self-duality of GS-sigma-models.
We study the type I and type II SCA's separately.

Comment: when talking about the grading, one usually use integer labeling, while the charges with respect to the U$(1)$ generators we will use ($D,B,\check R,\check \l$) are integer for the even generators and half integer for the odd generator (e.g there should really be a factor of 2 in figure \ref{fig:Dcharge}, and later in figure \ref{fig:typeIdecompositions}). In the next sections when we will write the charges with respect to the generators we will use integer numbers although they should be understood to be divided by 2, so at the end of the day when we will discuss the gradations with charges $\pm1,0$ only, these will really be the charges under the combination of the U$(1)$ generators.

\subsubsection{$\mathbb{Z}$-gradation of type I SCA's}
The type I SCA's include the su$(1,1|N\neq 2)$, su$(2,2|N\neq 4)$, psu$(1,1|2)$, psu$(2,2|4)$, osp$(2|2)\simeq$ su$(1,1|1)$, osp$(2|4)$.
Generally, the bosonic part of these SCA's is $\mathfrak{g}_{\bar 0}=\mathrm{so}(2,d)\oplus \mathrm{su}(N)\oplus \mathrm{u}(1)$.
We will refer to the last u$(1)$ as the hypercharge, which in the case of the 'psu' SCA's decouples from the SCA. The two 'osp' SCA's are missing the su$(N)$ subalgebra.

We will consider three $\mathrm{u}(1)$'s, generating consistent $\mathbb{Z}$-gradations of the SCA's, and their combinations which generates $\mathbb{Z}$-gradations with $|i|\leq 1$. We already considered $D$, which induces the following consistent $\mathbb{Z}$-gradation
\be
\mathfrak{g}_2=\mathrm{span}\{P\},\quad
\mathfrak{g}_1=\mathrm{span}\{Q,\hat Q\},\quad
\ee
$$
\mathfrak{g}_0=\mathrm{span}\{D,L\}\oplus \mathrm{u}(N),\quad
$$
$$
\mathfrak{g}_{-1}=\mathrm{span}\{S,\hat S\},\quad
\mathfrak{g}_{-2}=\mathrm{span}\{K\},
$$
and $B$ which induces the distinguished $\mathbb{Z}$-gradation
\be
\mathfrak{g}_1=\mathrm{span}\{Q,\hat S\},\quad
\mathfrak{g}_0=\mathrm{su}(M,M)\oplus \mathrm{u}(N),\quad
\mathfrak{g}_{-1}=\mathrm{span}\{\hat Q,S\}.
\ee
Lastly, we decompose the $\mathrm{su}(N)$ generators, $R^k_l$ with $k,l=1,...,N$ and $\sum_k R^k_k=0$, under $\mathrm{su}(N)\rightarrow \mathrm{s}(\mathrm{u}(P)\times \mathrm{u}(N-P))$. This divides the R-symmetry indices to $k,l,..=1,..,P$ and $k',l',...=P+1,...,N$, so the u$(P)$ generators are $R^k_l$ and the u$(N-P)$ generators are $R^{k'}_{l'}$, with the relation $\sum_k R^k_k+\sum_{k'} R^{k'}_{k'}=0$.
We define the generator, $\check R=\sum_k R^k_k-\sum_{k'} R^{k'}_{k'}$, inducing the $\mathbb{Z}$-gradation
\be
\mathfrak{g}_2=\mathrm{span}\{R^{k'}_l\},\quad
\mathfrak{g}_1=\mathrm{span}\{Q^k,\hat Q_{l'},S_{l'},\hat S^k\},\quad
\ee
$$
\mathfrak{g}_0=\mathrm{s}(\mathrm{u}(P)\times \mathrm{u}(N-P))\oplus \mathrm{su}(M,M),\quad
$$
$$
\mathfrak{g}_{-1}=\mathrm{span}\{Q^{k'},\hat Q_{l},S_{l},\hat S^{k'}\},\quad
\mathfrak{g}_{-2}=\mathrm{span}\{R^k_{l'}\}.
$$

Next we consider the combinations of $D,B$ and $\check R$ that give the $\mathbb{Z}$-gradations with $|i|\leq 1$ (up no normalization).
First we have the well known $U=D+B$ \cite{Beisert:2008iq}\cite{Berkovits:2008ic} which gives the non-consistent $\mathbb{Z}$-grading decomposition
\be\label{eq:typeID+B}
\mathfrak{g}=(P,Q)_1
      \oplus(L,D,\hat Q,\hat S,R)_0
      \oplus(K,S)_{-1}.
\ee
This decomposition holds for all the type I SCA's. In this case the invariant subalgebras are $\mathrm{(s)u}(M|N)$ for $\mathrm{(p)su}(2M|N)$, and $\mathrm{u}(1|2)$ for $\mathrm{osp}(2|4)$. Similarly for $U=D-B$ we get the same gradation with the hatted and unhatted generators interchanged.

Next, we consider $U=D+\check R$ which generates the decomposition
\be\label{eq:typeID+R}
\mathfrak{g}=(P,Q^{k},\hat Q_{k'},R_{k'}{}^l)_1
      \oplus(L,D,
      Q^{k'},\hat Q_{k},
      S_{k'},\hat S^{k},
      R_{k}{}^{l},R_{k'}{}^{l'})_0
      \oplus(K,S_k,\hat S^{k'},R_{k}{}^{l'})_{-1}
\ee
which was mentioned in \cite{Berkovits:2008ic} for psu$(2,2|4)$.
This decomposition cannot be applied for the 'osp' SCA's.
For this decomposition the invariant subalgebras are $\mathrm{(p)s}(\mathrm{u}(M|P)\times \mathrm{u}(M|N-P))$ for $\mathrm{(p)su}(2M|N)$.

Next, we consider $U=2B$ which generates the consistent distinguished-gradation
\be\label{eq:typeI2B}
\mathfrak{g}=(Q,\hat S)_1
      \oplus(P,K,D,L;R)_0
      \oplus(\hat Q,S)_{-1}.
\ee
This decomposition was not considered before in the context of T-duality, and implies the model may be self-dual under T-duality along fermionic directions only.

Lastly, we consider $U=B+\check R$ which generates the inconsistent-gradation
\be\label{eq:typeI2B+R}
\mathfrak{g}=(Q^{k},\hat S^{k},R_{k'}{}^l)_1
       \oplus (P,K,L,D,Q^{k'},\hat S^{k'},S_{k'},\hat Q_{k'},R_{k}{}^{l},R_{k'}{}^{l'})_0
       \oplus (S_k,\hat Q_k,R_{k}{}^{l'})_{-1}.
\ee
This decomposition also was not considered before in the context of T-duality. It is similar to the decomposition (\ref{eq:typeID+R}), interchanging the roles of the AdS and the sphere.

\begin{figure*}[!ht]
\centering
\epsfig{file=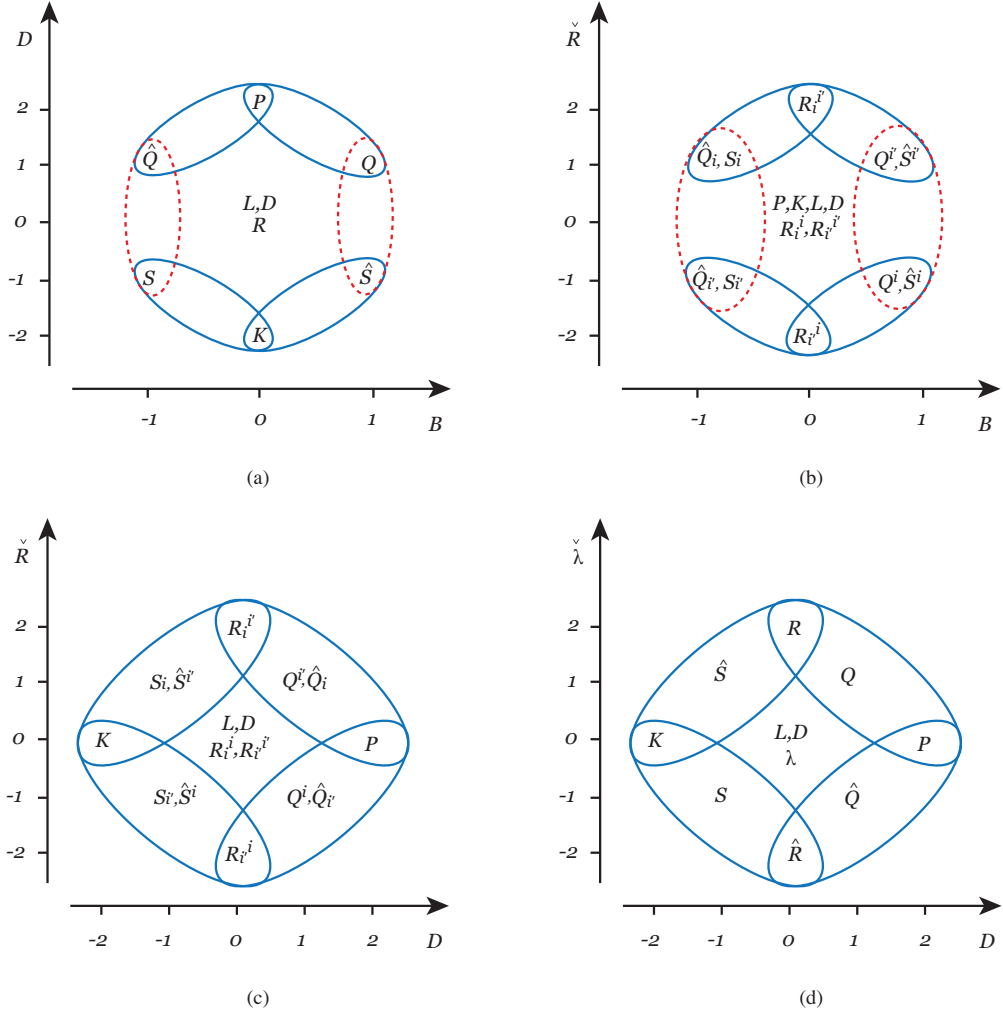,width=0.91\linewidth,clip=}
\caption{
$\mathbb{Z}$-gradation of type-I SCA's with $4N$-odd generators and R-symmetry $\mathrm{SU}(2M)\times \mathrm{U}(1)$. The abelian subalgebras are circled.
(a) Decomposition under $B$ and $D$. In this case the relevant $U(1)$'s are $(\pm)(D\pm B)$ (circled with solid blue contours) where the abelian subalgebra contains $d$-bosonic and $N$-fermionic generators, and $\pm 2B$ (circled with dashed red contours) where the abelian subalgebra contains $2N$-fermionic generators.
(b) Decomposition under $B$ and $\check R$. In this case the relevant $U(1)$'s are $(\pm)(\check R\pm B)$ (circled with solid blue contours) where the abelian subalgebra contains $M^2$-bosonic and $2M$-fermionic generators, and $\pm 2B$ (circled with dashed red contours) where the abelian subalgebra contains $2N$-fermionic generators.
(c) Decomposition under $D$ and $\check R$. In this case the relevant $U(1)$'s are $(\pm)(\check R\pm D)$ where the abelian subalgebra contains $d+M^2$-bosonic and $N$-fermionic generators.
(d) $\mathbb{Z}$-gradation of type-II SCA's with $4N$-odd generators and R-symmetry $R_1\oplus\l_0\oplus R_{-1}$. The abelian subalgebras are circled.
We have the decomposition under $\check \l$ and $D$. In this case the relevant $U(1)$'s are $(\pm)(D\pm \check \l)$ where the abelian subalgebra contains $d+\mathrm{dim}(R_1)$-bosonic and $N$-fermionic generators.
  }
\label{fig:typeIdecompositions}
\end{figure*}

We summarize the above decompositions in figure \ref{fig:typeIdecompositions} (a-c).
The results for $\mathrm{AdS}_5\times \mathrm{S}^5$ background are also given in table \ref{table:U1for-SCA}.
Note that in cases where the SCA is 'psu', the hypercharge is not a part of the SCA and so the $\mathbb{Z}$-gradation automorphisms are outer. The rest of the automorphisms are inner.
Decompositions with respect to combination, different from the ones presented by changing the relative signs of the generators are obvious and addressed in the figure.
One can find more decompositions of the SCA's, which we find less interesting with respect to the AdS backgrounds\footnote{For example, in the notation of $\mathrm{AdS}_5\times \mathrm{S}^5$ we have\\
$(P^{\a\dot\a},K^{\a\dot\a},L^{\a\a},L^{\dot\a\dot\a},Q^{i\a},\hat Q^{\dot\a}_{i'},S^{\a}_{i'},\hat S^{i\dot\a},R_{i'}{}^i)
      \oplus(P^{\bar\a\dot\a},P^{\a\bar{\dot\a}},K^{\bar\a\dot\a},K^{\a\bar{\dot\a}},L^{\a\bar\a},L^{\dot\a\dot\a},L^{\dot\a\bar{\dot\a}}, Q^{i\bar\a},Q^{i'\a},
      \hat Q^{\dot\a}_{i},\hat Q^{\bar{\dot\a}}_{i'},$ $S^{\a}_{i},S^{\bar\a}_{i'},\hat S^{i'\dot\a},\hat S^{i\bar{\dot\a}},R_{i}{}^i,R_{i'}{}^{i'})
      \oplus(K^{\bar\a\bar{\dot\a}},P^{\bar\a\bar{\dot\a}},L^{\bar\a\bar\a},L^{\bar{\dot\a}\bar{\dot\a}},S^{\bar\a}_i,\hat S^{i'\bar{\dot\a}},Q^{i'\bar\a},\hat Q^{\bar{\dot\a}}_i,R_{i}{}^{i'})
      $\\
   where $(\a,\dot\a)=(1,1),(1,2),(2,1),$ or $(2,2)$, $(\bar\a,\bar{\dot\a})$ take different value then $(\a,\dot\a)$, and $i=1,..,n\leq 4$, $i'=n+1,..,4$. The abelian subalgebra involves unphysical directions, $L$.
    The invariant subalgebra is  $\mathrm{ps}(\mathrm{u}(2|n)\times \mathrm{u}(2|4-n))$.}.

\subsubsection{$\mathbb{Z}$-gradation of type II SCA's}
The type II SCA's include the osp$(N\neq 2|2)$, osp$(N\neq 2|4)$, osp$(4^*|2)$, osp$(8^*|N)$, D$(2,1;\a)$, F$(4)$ and G$(3)$, but we'll consider only osp$(2N\neq 2|2)$, osp$(2N\neq 2|4)$, osp$(4^*|2)$, osp$(8^*|2N)$, D$(2,1;\a)$ and F$(4)$ which can be decomposed according to (\ref{eq:distinguishedII}) (the ones with even number of space-time supersymmetries).

As mentioned above the type II SCA's decompose under the charge assignment of $D$,
\be
\mathfrak{g}_2=\mathrm{span}\{P\},\quad
\mathfrak{g}_1=\mathrm{span}\{Q,\hat Q\},\quad
\ee
$$
\mathfrak{g}_0=\mathrm{span}\{D,L\}\oplus \mathrm{span}\{R,\l,\hat R\},\quad
$$
$$
\mathfrak{g}_{-1}=\mathrm{span}\{S,\hat S\},\quad
\mathfrak{g}_{-2}=\mathrm{span}\{K\}.
$$
and according to the gradation (\ref{eq:distinguishedII}), where the R-symmetry decomposes to $R_2\oplus\l_0\oplus\hat R_{-2}$ (the subscript indicates the gradation).
The R-symmetry decomposition for all type II SCA's with such decomposition is given in table \ref{table:Type II SCA's R-symmetry decomposition}.
The $\mathbb{Z}$-gradation of (\ref{eq:distinguishedII}) is induced by the generator $\check \l$ given in the table.

Combining the two, $U=D+\check \l$, we find that all type II SCA's have the inconsistent $\mathbb{Z}$-gradation
\be\label{eq:typeIID+lambda}
\mathfrak{g}=(P,Q,R)_1
      \oplus(L,D,\hat Q,\hat S,\l)_0
      \oplus(K,S,\hat R)_{-1}.
\ee
The invariant subalgebras of the decomposition are $\mathrm{u}(N|1)$, $\mathrm{u}(2|1)$, $u(1)\oplus \mathrm{osp}(2|4)$, $\mathrm{u}(2|N)$, $\mathrm{u}(N|2)$ and $\mathrm{u}(4|N)$ for
$\mathrm{osp}(2N|2)$, $\mathrm{D}(2,1;\a)$, $\mathrm{F}(4)$, $\mathrm{osp}(4^*|2N)$, $\mathrm{osp}(2N|4)$ and $\mathrm{osp}(8^*|2N)$ respectively

Note that all type II SCA's have no analog of the hypercharge, which assigns non-trivial charge to all odd generators and only to them.

\begin{table}[ht]
\caption{Type II SCA's R-symmetry decomposition.}
\centering
\scriptsize
\begin{tabular}{l | | l l l l }
\hline\hline
 SCA & R-symmetry & $[\l_0]$& $[R_1]$ & $\check \l$\\ [0.5ex]
\hline \hline
osp$(2N|2)$     & so$(2N)$                  & $\l^k_l$ $\in$ u$(N)$                         & $R_{kl}=-R_{lk}$  & $\sum_k\l^k_k$\\ \hline
osp$(2N|4)$     & so$(2N)$                  & $\l^k_l$ $\in$ u$(N)$                         & $R_{kl}=-R_{lk}$  & $\sum_k\l^k_k$\\ \hline
osp$(4^*|2N)$   & su$(2)\times$ usp$(2N)$   & ($\l^k_l$$\in$ u$(N))\times$ su$(2)$          & $R_{kl}=R_{lk}$   & $\sum_k\l^k_k$\\
{}              & {}                        & ($J_3$ $\in$ u$(1))\times$ usp$(2N)$          & $J_+$             & $J_3$\\ \hline
osp$(8^*|2N)$   & usp$(2N)$                 & $\l^k_l$ $\in$ u$(N)$                         & $R_{kl}=R_{lk}$   & $\sum_k\l^k_k$\\ \hline
D$(2,1;\a)$     & su$(2)\times$ su$(2)$     & $J_3$ $\in$ u$(1)$                            & $J_+$             & $J_3$\\ \hline
F$(4;0)$        & so$(7)$                   & $N_{ab}\times \l$ $\in$ so$(5)\times$ u$(1)$  & $R_a$             & $\l$\\ \hline
F$(4;2)$        & su$(2)$                   & $J_3$ $\in$ u$(1)$                            & $J_+$             & $J_3$\\ \hline
\end{tabular}
\begin{quote}
$[\l_0]$ and $[R_1]$ indicates the set of generators of the R-symmetry decomposed under $\check \l$ with charges 0 and 1 respectively.
The indices take the values: $k,l=1,...,N$ and $a=1,...,5$. We gave only $[R_1]$ where $[\hat R_{-1}]$ should be understood.
\end{quote}
\label{table:Type II SCA's R-symmetry decomposition}
\end{table}
\normalsize

We summarize the decomposition in figure \ref{fig:typeIdecompositions} (d).

\section{Green-Schwarz Sigma-models on Semi-Symmetric backgrounds}\label{sec:Green-Schwarz Sigma-models on Semi-Symmetric backgrounds}
\subsection{The action}
It is well known how to construct the Green-Schwarz sigma-model (GSSM) action on semi-symmetric spaces backgrounds $G/H$ with RR-flux, as was first done for the $\mathrm{AdS}_5\times \mathrm{S}^5$ background in \cite{Metsaev:1998it} (see e.g \cite{Adam:2007ws} for other backgrounds).
We introduce the left-invariant-one-form $\mathrm{j}=g^{-1}dg$ where $g\in G$, which take values in the SCA, and so decomposes under the $\mathbb{Z}_4$ automorphism to $\mathrm{j}=\mathrm{j}_0+\mathrm{j}_1+\mathrm{j}_2+\mathrm{j}_3$. We shall work in the $2d$-conformal basis where $\mathrm{j}=g^{-1}\p g$ and $\bar \mathrm{j}=g^{-1}\bar \p g$, so the GS sigma-model action is given by
\be\label{eq:GSaction}
S_{\mathrm{GS}}=\int d^2\s \Str\left(\mathrm{j}_2\bar \mathrm{j}_2+\frac{1}{2}\left(\mathrm{j}_1\bar \mathrm{j}_3-\mathrm{j}_3\bar \mathrm{j}_1\right)\right).
\ee
The action is invariant under left $G$-global multiplications of $g$ and right $H$-local multiplications. It is also invariant under the local fermionic $\k$-symmetry transformation, where the rank of the transformation depends on the coset background \cite{Metsaev:1998it}\cite{Zarembo:2010sg}.
\subsection{Integrability}
The sigma-model is known to be integrable \cite{Bena:2003wd} by introducing a flat-connection depending on a spectral parameter.
In the present paper we shall work with the flat connection
\be\label{eq:flatconnection}
A(z)=\mathrm{j}_{(0)}
+z \mathrm{j}_{(1)}
+\frac{1}{2}(z^2+z^{-2}) \mathrm{j}_{(2)}
+z^{-1} \mathrm{j}_{(3)}
-\frac{1}{2}(z^2-z^{-2}) \ast \mathrm{j}_{(2)}
\ee
where $z\in \mathbb{C}$ is the spectral parameter. Noting that
\be\label{eq:jBjF}
\mathrm{j}_{(0)}=\frac{1}{2}(1+\O)\mathrm{j}_B,\quad
\mathrm{j}_{(2)}=\frac{1}{2}(1-\O)\mathrm{j}_B,\quad
\ee
$$
\mathrm{j}_{(1)}=\frac{1}{2}(1-i\O)\mathrm{j}_F,\quad
\mathrm{j}_{(3)}=\frac{1}{2}(1+i\O)\mathrm{j}_F,
$$
where $\mathrm{j}_B$ and $\mathrm{j}_F$ are the even and odd parts of the current $\mathrm{j}$, we can rewrite the flat-connection
\be
A(z)=
 \frac{1}{4}(z+z^{-1})^2 \mathrm{j}_B
-\frac{1}{4}(z-z^{-1})^2 \O(\mathrm{j}_B)
-\frac{1}{4}(z^2-z^{-2}) \ast (\mathrm{j}_B-\O(\mathrm{j}_B))
\ee
$$
+\frac{1}{2}(z+z^{-1})\mathrm{j}_F
-\frac{i}{2}(z-z^{-1}) \O(\mathrm{j}_F).
$$

Once we have the flat-connection we can construct an infinite tower of conserved charges using the monodromy matrix,
which is given by
\be
M(z)=\overrightarrow{\mathcal{P}\exp}\int_\g a(z)
\ee
where $\mathcal{P}$ stands for path ordering and
and $a(z)$ is the gauge invariant flat-connection, related to $A(z)$ by $d+a(z)=g(d+A(z))g^{-1}$.
Expanding $M(z)$ around $z=\pm 1$, one can extract the conserved n-local charges.

\section{T-duality of the GS Sigma-Models}\label{sec:T-duality of the GS Sigma-Models}

\subsection{Assumptions}
In this section we summarize our assumptions and motivate them.
First of all, we T-dualize along all directions associated with the generators of $A_1$ (see section \ref{sec:Z-gradation}).
This means we T-dualize along the coordinates that couple to the generators in $A_1$ in the expression for $j=g^{-1}dg$ when expanded to first order in all coordinates.
Thus,  these coordinates should appear in the action only through their derivatives, in order to have translation isometry.
For this reason we parameterize the group element representative as
\be\label{eq:parametrization2}
g(x,y)=e^{a(x)}e^{b(y)},
\ee
with $a(x)=x^I L_I\in A_1$ and assume $b(y)=y^\a L_\a\in\!\!\!\!\!/ ~A_1$.

As will be shown in the next subsection the WZW piece of the Lagrangian is given by
$L_{WZW}=-\frac{i}{2}\Str(\mathrm{j}_F\O(\bar \mathrm{j}_F))$.
To zeroth order in the fermions, keeping only their derivatives, this is the term that should produce the quadratic term for the fermionic coordinates.
In order to perform the T-duality transformation this term must be non-singular, namely the matrix that couples the fermions which we want to T-dualize along, should be invertible.
For type I SCA's, when we want to T-dualize, for example, along span$\{P,Q\}$ directions, it means that in order to have a non-singular coupling for the fermionic coordinates, $\Str(Q\O(Q))$ must be non-trivial, namely $\O(Q)\sim S$, which is the case if (\ref{eq:Z4Z3condition}) is satisfied.

This condition also implies the existence of a non-singular fermionic quadratic term for type II SCA's.
But for type II, in principle, one can also have a non-singular quadratic term also if (\ref{eq:Z4Z3condition}) is not satisfied, by conjugating $Q$ with $g_{\hat R}=\exp(z\cdot \hat R)$, so $g^{-1}_{\hat R}Q g_{\hat R}\sim Q+z\hat Q$. This means that if $\O(\hat Q)\sim S$, $\Str(Q\O(\hat Q))$ is not trivial. As found in \cite{Adam:2009kt}, the coupling can still be singular if, for example, the dimension of the representation of $\hat R$ is odd.
We note that this automorphism relation implies $\O(R)\sim R$ and $\O(\hat R)\sim \hat R$,
while the non-trivial bilinear form for these generators is of the form $B(R,\hat R)$, so if we want also to T-dualize along the $R$ direction (as we do according to the first assumption) we will have to parameterize $g\sim e^{yR}e^{z\hat R}$. This will give the kinetic term
$L_{\mathrm{kin}}\sim \p y(z^2\bar\p y+\bar\p z)+\mathrm{c.c}$ which after T-duality along $y$ gives $\tilde L_{\mathrm{kin}}\sim\frac{\p \tilde y\bar\p \tilde y}{z^2}+dy\wedge dz$.
Thus, we do not get back the same background, since we did not have this B-field in the original action.
For these reasons we impose that (\ref{eq:Z4Z3condition}) is satisfied.

We note that if the parametrization includes terms in $A_{-1}$, it is unlikely to get self-duality, where again T-duality will produce a B-field instead of part of the metric. For example, if we take $g\sim e^{xP}e^{yK}$, then $j=\p x+\p y +...$ (where the ellipsis stands for higher powers of $y$ which will not affect the point we are making), and the kinetic term is $L_{\mathrm{kin}}\sim \p x\bar\p y+\bar\p x\p y-\p x\bar\p x-\p y\bar\p y$. Under T-duality along $x$, $\tilde L_{\mathrm{kin}}\sim dx\wedge dy-\p x\bar\p x-\p y\bar\p y$. So the background is not invariant under this transformation. This motivates us to consider only parameterizations of the form
\be\label{eq:parametrization}
g(x,y)=e^{a(x)}e^{b(y)},
\ee
with $a(x)=x^I L_I\in A_1$ and $b(y)=y^\a L_\a\in B_0$.
The indices $I,J,...$ will denote generators in $A_1$. The indices $\a,\b,...$ will denote generators in $B_0$ which are eigenstates of $\O(\cdot)$.
The current decomposes to
\be\label{eq:currentdecomposition}
\mathrm{j}=g^{-1}d g=e^{-b(y)}dx^I L_Ie^{b(y)}+e^{-b(y)}de^{b(y)}\equiv J(x,y)+j(y)
\ee
where $J\in A_1$ and $j\in B_0$.
In order to parameterize $g$ as in (\ref{eq:parametrization}), we will have to use the local $H$-gauge-symmetry and $\k$-symmetry.
As we will see later the problem will reduce to computing the rank of the $\k$-symmetry. For the $\mathbb{Z}$-gradations introduced in section \ref{sec:Z-gradation}, we have to gauge away at least quarter of the odd degrees of freedom and in one case at least half.
The decomposition of the current (\ref{eq:currentdecomposition}) implies that the action will not have mixed terms of $J$ and $j$.

To summarize, we assume the backgrounds to satisfy (\ref{eq:Z4Z3condition}) and that we can parameterize the action as in (\ref{eq:parametrization}).

\subsection{T-duality of the Green-Schwarz sigma-model}
We rewrite the action (\ref{eq:GSaction}) in terms of $\mathrm{j}_B$ and $\mathrm{j}_F$ defined in (\ref{eq:jBjF}).
First we consider the WZW term (assuming $A_1$ containing odd generators)
\be
L_{\mathrm{WZW}}=\frac{1}{2}\Str(\mathrm{j}_{1}\bar \mathrm{j}_{3}-\mathrm{j}_{3}\bar \mathrm{j}_{1})
=\frac{1}{8}\Str((1+i\O)\mathrm{j}_{F}(1-i\O)\bar \mathrm{j}_{F}-(1-i\O)\mathrm{j}_{F}(1+i\O)\bar \mathrm{j}_{F})
\ee
$$
=-\frac{i}{2}\Str(J_F\O(\bar J_F)+j_F\O(\bar j_F)).
$$
It is natural to define a bilinear-form\footnote{
This Bilinear form is consistent ($B(X,Y)=0$ $\forall$ $X\in \mathfrak{g}_{\bar 0}$ and $Y\in \mathfrak{g}_{\bar 1}$), but neither supersymmetric ($B(X,Y)=(-)^{|X|+|Y|}B(Y,X)$) nor invariant ($B([X,Y\},Z)=B(X,[Y,Z\})$), see \cite{Kac:1977em}\cite{Frappat:1996pb}.} $\eta_{AB}=\Str(L_A \O(L_B))$, $L_A, L_B\in \mathfrak{g}$ and its inverse satisfying $\eta^{AB}\eta_{BC}=\d^A_C$.
We will raise and lower indices using the bilinear-form.
Note that it satisfies $\eta_{AB}=\eta_{BA}$ for both bosonic and fermionic parts\footnote{This is a bit unusual property due to our special definition of the bilinear-form, which further implies $\psi^A\psi^B\eta_{AB}=0$ if $\psi$ is odd. Usually the fermionic part of the bilinear-form is anti-symmetric.}.
We further introduce the notation $\mathrm{j}=\mathrm{j}^A L_A=\eta^{AB} \Str(\mathrm{j}  \O(L_B)) L_A$.
Returning to the WZW term we find
\be
L_{\mathrm{WZW}}=\frac{i}{2}J_F^I\bar J_F^J \eta^F_{IJ}
+\frac{i}{2}j_F^\a \bar j_F^\b \eta^{(1,3)}_{\a\b},
\ee
where $\eta^{(1,3)}_{\a\b}$ means that $\a\in \mathcal{H}_1$ and $\b\in \mathcal{H}_3$ or $\a\in \mathcal{H}_3$ and $\b\in \mathcal{H}_1$.

Next, we consider the kinetic part of the GS action
\be
L_{\mathrm{kin}}=\Str(\mathrm{j}_2\bar \mathrm{j}_2)
=\frac{1}{4}\Str((1-\O)\mathrm{j}_B(1-\O)\bar \mathrm{j}_B)
\ee
$$
=\frac{1}{2}\Str(J_B\bar J_B-J_B\O(\bar J_B)+j_B\bar j_B-j_B\O(\bar j_B)).
$$
Since we work with $\O(A_1)\in  A_{-1}$, the first term vanishes and we get
\be
L_{\mathrm{kin}}=\frac{1}{2}\Str(-J_B^I\bar J_B^J L_I\O(L_J)+j_B^\a\bar j_B^\b L_\a(1-\O)(L_\b))
\ee
$$
=-\frac{1}{2}J_B^I\bar J_B^J \eta^B_{IJ}-j_B^\a\bar j_B^\b \eta^{(2)}_{\a\b},
$$
where the superscript in $\eta^{(2)}$ reminds us that we take only generators in $\mathcal{H}_2$.

All in all we find the GS sigma-model Lagrangian is given by
\be
L_{\mathrm{GS}}
=-\frac{1}{2}J_B^I\bar J_B^J \eta^B_{IJ}-j_B^\a\bar j_B^\b \eta^{(2)}_{\a\b}
+\frac{i}{2}J_F^I\bar J_F^J \eta^F_{IJ}
+\frac{i}{2}j_F^\a \bar j_F^\b \eta^{(1,3)}_{\a\b}.
\ee
We want to T-dualize the sigma-model along the directions $x^I$ which appear only through their derivatives in $J$, so we introduce the gauge fields $A=\p x^I L_I$ and $\bar A=\bar \p x^I L_I$, and add the Lagrange multiplier term
\be
L_{\mathrm{LM}}=-\frac{1}{2}\p\tilde x^I\Str(e^b \bar A' e^{-b}\O(L_I))
+\frac{1}{2}\bar\p\tilde x^I\Str(e^b A' e^{-b}\O(L_I))
\ee
where $A'=e^{-b} A e^{b}=A'_B+A'_F\in A_1$ (note that the index $I$ runs over both bosonic and fermionic generators of $A_1$).
We define
\be
W_J=\Str(e^{-b}\p\tilde x^I \O(L_I)e^{b} L_J)=\Str(e^{-\O(b)}\p\tilde x^I \O^2(L_I) e^{\O(b)} \O(L_J)),
\ee
and rewrite
\be
L_{\mathrm{LM}}=\frac{1}{2}A'^I \bar W_I-\frac{1}{2}\bar A'^I W_I.
\ee
The EOM for $J$ give
\be
\bar A'^J_B \eta^B_{IJ}=\bar W_I,\quad
A'^I_B \eta^B_{IJ}=-W_J,\quad
\ee
$$
\bar A'^J_F \eta^F_{IJ}=i \bar W_I,\quad
A'^I_F \eta^F_{IJ}=i W_J,
$$
or by using the metric definitions
\be\label{eq:dualcurrent}
\bar A'^I_B =\bar W^I,\quad
A'^I_B =-W^I,\quad
\ee
$$
\bar A'^I_F =i \bar W^I,\quad
A'^I_F =i W^I.
$$
Plugging these into the action we get
\be
\tilde L_{GS}
=-\frac{1}{2}W_B^I\bar W_B^J \eta^B_{IJ}-j_B^\a\bar j_B^\b \eta^{(2)}_{\a\b}
+\frac{i}{2}W_F^I\bar W_F^J \eta^F_{IJ}
+\frac{i}{2}j_F^\a \bar j_F^\b \eta^{(1,3)}_{\a\b}.
\ee

We can also construct $\tilde L_{\mathrm{GS}}$, if instead of the original parametrization we take $\tilde g=e^{\O(a)}e^b$ and replace $x$ with $\tilde x$, so $\tilde j=j$ is unaffected as desired, and
\be
\tilde J\equiv \tilde J^K \O(L_K)=\Str(e^{-b}\p\tilde x^J\O(L_J) e^b \O^2(L^K))\O(L_K)
\ee
so
\be
\tilde J_K=\Str(e^{-b}\p\tilde x^J\O(L_J) e^b \O^2(L_K))=(-)^K\Str(e^{-b}\p\tilde x^J\O(L_J) e^b L_K)=(-)^K W_K.
\ee
Note that the minus factor $(-)^K$ doesn't affect the action since the $\tilde J$'s appear quadratically.
We also used $\Str(\O(L_I)\O^2(L_J))=\Str(L_I \O(L_J))$ in the $W\bar W$ part of the WZW term.
These two parametrization are related by the automorphism $\O$ if we redefine the coordinates in $e^b$.
If $b=y^\a L_\a$, then the dual coordinates should be $\tilde y^\a=i^{\a}y^{\a}$ (where the $\a$ in $i^\a$ indicates the $\mathbb{Z}_4$ grading of $L_\a$), so
\be
\O(g(x,y))=e^{\O(a(x))}e^{\O(b(y))}=e^{\O(a(\tilde x))}e^{b(\tilde y)}=\tilde g(\tilde x,\tilde y).
\ee
We also note that $\O(\tilde b)=\O(\tilde y^\a L_\a)=y^\a L_\a=b$, so (from now on a tilde over a current means we take the original current and plug the dual coordinates)
\be
\tilde J_K\equiv J_K(\tilde x,\tilde y)=\Str(e^{-\tilde b}\p\tilde x^I L_I e^{\tilde b} \O(L_K))
\ee
$$
=\Str(e^{-b}\p\tilde x^I \O(L_I) e^{b} \O^2(L_K))
=(-)^K W_K.
$$
Similarly
\be
\tilde j_{L_\a}
\equiv \tilde j^\a L_\a
\equiv j^\a(\tilde y) L_\a
=\Str(e^{-\tilde b}\p e^{\tilde b} \O(L^\a))L_\a
\ee
$$
=i^{-\a}\Str(e^{-b}\p e^{b} \O^2(L^\a))\O(L_\a)
=\O(j_{L_\a}).
$$
To summarize, using (\ref{eq:dualcurrent}) we find that the left-invariant-one-form transforms as
\be\label{eq:currenttransformation}
\begin{array}{l}
\tilde J_B=\ast J_B,\\
\tilde J_F=i J_F,\\
\tilde j_{L_\a}=\O(j_{L_\a}).
\end{array}
\ee

\subsection{Flat-connection transformation under T-self-duality }
Now that we know how the left-invariant-one-form transforms under T-duality, see (\ref{eq:currenttransformation}), we can find the general transformation of the flat-connection (\ref{eq:flatconnection}).
For simplicity of notation we define $b(z)=(z+z^{-1})/2$, so the flat connection takes the form
\be\label{eq:flatconnection2}
A(z)=
b(z)^2 \mathrm{j}_B
+b(iz)^2 \O(\mathrm{j}_B)
+ib(z)b(iz)\ast (\mathrm{j}_B-\O(\mathrm{j}_B))
+b(z)\mathrm{j}_F
-b(iz) \O(\mathrm{j}_F).
\ee
$$
=b(z)^2 (J+j)_B
+b(iz)^2 \O((J+j)_B)
+ib(z)b(iz)\ast ((J+j)_B-\O((J+j)_B))
$$
$$
+b(z)(J+j)_F
-b(iz) \O((J+j)_F).
$$
Let now write down the dual flat-connection, plugging (\ref{eq:currenttransformation}) into (\ref{eq:flatconnection2})
\be
\tilde A(z)=
b(z)^2 (\ast J_B+\O(j_B))
+b(iz)^2 (\O(\ast J_B)+j_B)
+ib(z)b(iz)(J_B+\ast \O(j_B)-\O(J_B)-\ast j_B)
\ee
$$
+b(z)(iJ_F+\O(j_F))
-b(iz) (\O(iJ_F)-j_F).
$$
As in \cite{Beisert:2008iq}, one can relate the two flat-connection by using a $z$-dependent automorphism $U_z(\cdot)$ which acts as follows
\be
U_z(J)=f(z)\O(J),\quad
U_z(\O(J))=(-)^J f^{-1}(z)J,\quad
\ee
$$
U_z(j)=\O(J),\quad
U_z(\O(j))=(-)^j j,
$$
with $f(z)=-ib(iz)/b(z)$, so
\be
U_z(A(z))=\tilde A(z).
\ee
This automorphism is a composition of the $\mathbb{Z}_4$ automorphism and the one-parameter automorphism induced by the $\mathbb{Z}$-gradation (\ref{eq:Zgradationautomorphism}), with $\l=f(z)$. That is
\be
U_z(\cdot)=\s_{f(z)}(\O(\cdot)).
\ee
The U$(1)$ used in \cite{Beisert:2008iq} is $U=D+B$ where $B$ is the hypercharge. In table \ref{table:U1for-SCA} we give more examples of possible U$(1)$ charges for the GS sigma-model on AdS$_5\times $S$^5$, as explained in section \ref{sec:Z-gradation}.
\begin{table}[ht]
\caption{U(1) charges.}
\centering
\scriptsize
\begin{tabular}{c | | c c c c c c c c}
\hline\hline
 U$(1)$ & $P$ &  $Q^{i\a}$ & $Q^{i'\a}$ & $\hat Q^{\dot\a}_{i}$ & $\hat Q^{\dot\a}_{i'}$ & $R^{i}_{i'}$ & $R^{i}_{i}$ & D\\ [0.5ex]
\hline \hline
$D$ & 1 & 1/2 & 1/2 & 1/2 & 1/2 & 0 & 0 & 0\\ \hline
$B$ & 0 & 1/2 & 1/2 & -1/2 & -1/2 & 0 & 0 & 0\\ \hline
$\check R$ & 0 & 1/2 & -1/2 & -1/2 & 1/2 & 1 & 0 & 0\\ \hline
\hline
$D+B$ & 1 & 1 & 1 & 0 & 0 & 0 & 0 & 0\\ \hline
$D-B$ & 1 & 0 & 0 & 1 & 1 & 0 & 0 & 0\\ \hline
$D+\check R$ & 1 & 1 & 0 & 0 & 1 & 1 & 0 & 0\\ \hline
$D-\check R$ & 1 & 0 & 1 & 1 & 0 & -1 & 0 & 0\\ \hline
\hline
$B+\check R$ & 0 & 1 & 0 & -1 & 0 & 1 & 0 & 0\\ \hline
$B-\check R$ & 0 & 0 & 1 & 0 & -1 & -1 & 0 & 0\\ \hline
$2B$ & 0 & 1 & 1 & -1 & -1 & 0 & 0 & 0\\ \hline
\end{tabular}
\begin{quote}
We give U$(1)$ generators, $D,B$ and $\check R$, with respect to gradations of the SCA psu$(2,2|4)$.
The charge of the SCA's generators under the U$(1)$'s and under their combinations that decomposes the SCA to $A_1\oplus B_0\oplus A_{-1}$ are given.
$i$ and $i'$ are the su$(4)$ indices when broken to $\mathrm{su}(2)\times \mathrm{su}(2)$.
The charges of the generators related to those in the table by $\O$, have the opposite charge, while the rest of the generators not given in the table have zero charge.
\end{quote}
\label{table:U1for-SCA}
\end{table}
\normalsize

Repeating the arguments of \cite{Beisert:2008iq}\cite{Beisert:2009cs}\cite{Berkovits:2008ic}, the $A_1$ Noether charges becomes trivial, the $A_{-1}$ charges gets lifted and become non-local, and the $B_0$ generators remains local and transforms into themselves up to commutators and boundary terms.

\subsection{Quantum consistency of the T-Self-Duality transformation}
In previous subsections we have shown that the GSSM is self-dual under T-duality where we made some assumption regarding the $\mathbb{Z}$-gradation and $\mathbb{Z}_4$ automorphisms of the SCA, and the possibility to kappa-gauge fix the action in a certain way.
Though it is enough at the level of the classical action, at the quantum level we should also worry about the dilaton transformation under the T-duality transformation \cite{Buscher:1987qj}\cite{Buscher:1987sk}. In order to have self-duality at the quantum level we need the dilaton to be left invariant under the transformation \cite{Berkovits:2008ic}. This means that the super-Jacobian of the transformation should equal one. This depends on how the generators in $A_1$ transform under conjugation with $e^b$.

For the $\mathbb{Z}$-gradation induced by $B+D$ and $D+\check \l$ for the type I and II respectively,
a necessary condition for invariance of the dilaton is that the number of $Q$'s will be twice the number of $P$'s, since their charge under $D$ is half of the charge of the $P$'s, where $D\in \mathcal{H}_2$ and so should appears in the parametrization as in \cite{Berkovits:2008ic}. For type II, the number of $Q$'s should also be twice the number of $R$'s for the same reason (but with respect to $\check \l\in \mathcal{H}_2$), namely the number of $P$'s should also equal the number of $R$'s.
We find that all SCA's satisfying this condition have vanishing Killing-form.
It is known that GSSM on backgrounds based on supergroups with vanishing Killing-form are special, e.g these models are conformal invariant at one-loop \cite{Berkovits:1999zq}\cite{Kagan:2005wt}\cite{Zarembo:2010sg}.
This means that self-duality might be related to conformal invariance of the sigma-model.
This condition is also required by the other $\mathbb{Z}$-gradations considered in section \ref{sec:Z-gradation}. For example when $U=2B$, if the supergroup does not have vanishing Killing-form, the hypercharge - $B\in \mathfrak{g}_{\bar 0}$ and also $B\in \mathcal{H}_2$ so we can't gauge it away while all generators in $A_1$ (span$\{Q,\hat S\}$) have the same charge under it.
As we will see later, there are cases where the action is self-dual classically, but the dilaton transforms non-trivially.

\section{Classification of the backgrounds}\label{sec:Classification of the backgrounds}
In this section we would like to find all the GSSM on semi-symmetric backgrounds which are self-dual under T-duality along the directions of the abelian-subalgebras found in section \ref{sec:Z-gradation}.
In order to do so, we have to find the backgrounds based on SCA's and $\O$'s satisfying:
\begin{itemize}\label{conditions}
  \item $\O(U)=-U$.
  \item Rank($\k$-symmetry) $\geq$ dim($\mathfrak{g}_{\bar 1}$)/4.
  \item The Killing-form is degenerate.
\end{itemize}
We discuss separately the type I and type II SCA's.
We consider backgrounds of dimension $\leq 10$ with an AdS bosonic subspace.
For AdS$_{n>2}$ we T-dualize along even number of bosonic directions or else we will switch type IIA with type IIB and vice versa.

For later use we note that the $\mathbb{Z}_4$'s satisfying (\ref{eq:Z4Z3condition}), act on the SCA in the same way, namely
\be\label{eq:Z4DIA}
\O(P)\sim K,\quad
\O(D)\sim D,\quad
\O(L)\sim L,\quad
\ee
$$
\O(Q)\sim S,\quad
\O(\hat Q)\sim \hat S,
$$
and for the type I R-symmetry
$$
\O(R)\sim R,
$$
and for type II
$$
\O(R)\sim \hat R,\quad
\O(\l)\sim \l.
$$

\subsection{Type I}
In table \ref{table:Type_I_QPAdS} we give the $\mathbb{Z}_4$ automorphisms which satisfy the condition (\ref{eq:Z4Z3condition}) for AdS-semi-symmetric spaces.
As one can see, such a $\mathbb{Z}_4$ automorphism does not always exists. For each semi-symmetric space induced by $\O$, we give the rank of the kappa-symmetry, which should be $\geq$ then quarter of the number of odd generators of the SCA (or half of then in case where $A_1=$span$\{Q,\hat S\}$)\footnote{Note that whenever the rank of kappa-symmetry is large enough to eliminate quarter of the odd degrees of freedom, it actually large enough to eliminate half of them. This can expected since one can interpret the charges transformation under the T-duality as a sort of rotation in the Yangian space, and in some sense the $(Q,\hat S)$ duality can be thought of as a composition of $(P,Q)$ and $(K,\hat S)$, so these two exist then we expect the other to also exist.}. We also write if the super-Jacobian is unity or not based on the degeneracy of the Killing-form. By these criteria we determine whether the sigma-model is self-dual or not at the classical and quantum levels.

In table \ref{table:Type_I_QPnonAdS} we give the $\mathbb{Z}_4$ automorphisms which satisfy (\ref{eq:Z4Z3condition}) but induce semi-symmetric space which in not a product of AdS background with some other compact space.
In all of these cases the rank of kappa-symmetry is zero.

To summarize, we find there are two models, the $\mathrm{AdS}_5\times \mathrm{S}^5$ and $\mathrm{AdS}_2\times \mathrm{S}^2$ which are self-dual at the quantum level, these results were proven in \cite{Berkovits:2008ic}\cite{Beisert:2008iq} and in \cite{Adam:2009kt} respectively.
There are two more backgrounds which are self-dual only at the classical level, these are the $\mathrm{AdS}_5\times \mathrm{S}^1$ and $\mathrm{AdS}_2\times \mathrm{S}^1$, the first one was given in \cite{Hao:2009hw} (although the dilaton shift was not discussed) and the second is new.
These models should not be conformal invariant by themselves and one have to add D-branes in order to make them conformal invariant \cite{Klebanov:2004ya}. It might be that after adding open strings degrees of freedom one would find quantum consistent self-duality.
\begin{table}[ht]
\caption{AdS Semi-symmetric spaces based on type-I SCA's.}
\centering
\scriptsize
\begin{tabular}{c | | c c c c c c}
\hline\hline
  & PSU$(1,1|2)$ & $\begin{array}{c}
                     \mathrm{SU}(1,1|N) \\
                     N > 2
                   \end{array}$
   & PSU$(2,2|4)$ & $\begin{array}{c}
                     \mathrm{SU}(2,2|2N) \\
                     N\neq 2
                   \end{array}$ & OSP$(2|2)$ & OSP$(2|4)$\\ [0.5ex]
\hline \hline
$B_0$ & SU$(1|2)$ & U$(1|N)$ & SU$(2|4)$ & U$(2|2N)$ & U$(1|1)$ & U$(1|2)$ \\
\hline
$\begin{array}{c}
  \mathrm{Invariant }\\
  \mathrm{sub-alg}.
\end{array}$
  & U$(1)^2$ & $\mathrm{SO}(1,1) \times \mathrm{SO}(N)$
   & USp$(2,2)\times$ USp$(4)$ & $\mathrm{USp}(2,2)\times \mathrm{USp}(2N)$
                & U$(1)$ & $\varnothing$\\
   \hline
$\begin{array}{c}
  \mathrm{Bosonic-}\\
  \mathrm{subspace}
\end{array}$
  & AdS$_2\times$ S$^2$ & $\mathrm{AdS}_2\times \mathrm{S}^1 \times \mathrm{AI}(N)$ &
               AdS$_5\times$ S$^5$ &
               $\mathrm{AdS}_5\times \mathrm{S}^1 \times \mathrm{AII}(N)$
                & AdS$_2\times$ S$^1$ & $\varnothing$\\
               \hline
$\begin{array}{c}
  \mathrm{Rank~of}\\
  \k-\mathrm{symm}.
\end{array}$
  & 4 & 0 & 16 & $\begin{array}{c}8~\mathrm{for}~$N=1$,\\ \mathrm{else}~0\end{array}$ & 2 & $\varnothing$\\
  \hline
sJacobian
  & 1 & $\varnothing$ & 1 & $\neq~1$ & $\neq~1$ & $\varnothing$\\
  \hline
Self-dual
  & Yes-Q & No & Yes-Q & $\begin{array}{c}\mathrm{Yes-C~for}~$N=1$,\\ \mathrm{else~No}\end{array}$ & Yes-C & No\\   [1ex]
\hline
\end{tabular}
\begin{quote}
When the bosonic symmetric space is not AdS or a sphere, we symbolize it according to the Cartan classification, see \cite{Helgason:2001}. Note that dim$(\mathrm{AI}(N))={\frac{(N-1)(N+2)}{2}}$ and dim$(\mathrm{AII}(N))=(N-1)(2N+1)$.
We symbolize with $\varnothing$ the cases where we don't have relevant $\mathbb{Z}_4$ automorphism generating AdS space, and when the calculation of the super-Jacobian is not relevant. Yes-Q means the model is self-dual at the quantum level, and Yes-C means the model is self-dual only at the classical level. Note that SU$(1,1|1)\simeq$OSP$(2|2)$.
\end{quote}
\label{table:Type_I_QPAdS}
\end{table}
\normalsize\\
\begin{table}[ht]
\caption{Non-AdS Semi-symmetric spaces based on type-I SCA's.}
\centering
\scriptsize
\begin{tabular}{c | | c c c c c c}
\hline\hline
  & PSU$(1,1|2)$ & $\begin{array}{c}
                     \mathrm{SU}(1,1|N) \\
                     N\neq 2
                   \end{array}$
   & PSU$(2,2|4)$ & $\begin{array}{c}
                     \mathrm{SU}(2,2|N) \\
                     N\neq 4
                   \end{array}$ & OSP$(2|2)$ & OSP$(2|4)$\\ [0.5ex]
\hline \hline
$B_0$ & SU$(1|2)$ & U$(1|N)$ & SU$(2|4)$ & U$(2|N)$ & U$(1|1)$ & U$(1|2)$ \\
\hline
$\begin{array}{c}
  \mathrm{Invariant }\\
  \mathrm{sub-alg}.
\end{array}$
  & $\varnothing$ & $\varnothing$
   & SO$(2,2)\times$ SO$(4)$ & $\mathrm{SO}(2,2)\times \mathrm{SO}(N)$
                & $\varnothing$ & $\mathrm{U}(2)$\\
   \hline
$\begin{array}{c}
  \mathrm{Bosonic-}\\
  \mathrm{subspace}
\end{array}$
  & $\varnothing$ & $\varnothing$ &
               $\mathrm{AI}(2,2)\times \mathrm{AI}(4)$ &
               $\mathrm{AI}(2,2)\times \mathrm{AI}(N)\times \mathrm{S}^1$
                & $\varnothing$ & $\mathrm{CI}(4)\times \mathrm{S}^1$\\
               \hline
$\begin{array}{c}
  \mathrm{Rank~of}\\
  \k-\mathrm{symm}.
\end{array}$
  & $\varnothing$ & $\varnothing$ & 0 & 0 & $\varnothing$ & 0\\
  \hline
Self-dual
  & No & No & No & No & No & No\\   [1ex]
\hline
\end{tabular}
\begin{quote}
The notations are the same as in table \ref{table:Type_I_QPnonAdS}. Note that dim$(\mathrm{AI}(N))={\frac{(N-1)(N+2)}{2}}$ and dim$(\mathrm{C}(4))=6$.
We symbolize with $\varnothing$ the cases where we don't have a $\mathbb{Z}_4$ automorphism generating non-AdS space.
\end{quote}
\label{table:Type_I_QPnonAdS}
\end{table}
\normalsize

Finally, we note that the background AdS$_2\times \mathbb{CP}^n$ has been claimed to be self-dual
 in \cite{Hao:2009hw}. This background does not appear in our classification as self-dual.
 The calculation in \cite{Hao:2009hw} has a flaw: the authors redefine the odd generators below (eq. 27) in that paper, but the redefinition is not one-to-one.
  Thus, the generators are no longer independent, and the algebra does not close and does not represent SU$(1,1|N)$.  The coset space is therefore not a quotient of a super-Lie-algebra and of-course does not give the AdS$_2\times \mathbb{CP}^n$ background.

\subsection{Type II}
In table \ref{table:Type_II_QPAdS2} we give the $\mathbb{Z}_4$ automorphisms which satisfy the condition (\ref{eq:Z4Z3condition}) and gives AdS$_2\times \mathcal{M}$ backgrounds, and in table \ref{table:Type_II_QPAdSn} backgrounds with AdS$_{n>3}\times \mathcal{M}$. The notations are the same as for the type I SCA's.

In table \ref{table:Type_II_QPnonAdS} we give the $\mathbb{Z}_4$ automorphisms which satisfy (\ref{eq:Z4Z3condition}) but induce semi-symmetric space which in not a product of AdS background with some other compact space.

For the F$(4;0)$ SCA, the semi-symmetric space satisfying (\ref{eq:Z4Z3condition}) is $\mathrm{AdS}_2\times \mathrm{BDI}(3;4)$ with the invariant subalgebra $\mathrm{SO}(3)\times \mathrm{SO}(4)\times \mathrm{U}(1)$ with 14-dimensional space-time so we don't treat it in table \ref{table:Type_II_QPAdS2}. Similarly we omit F$(4;2)$ from table \ref{table:Type_II_QPnonAdS} were the relevant semi-symmetric space is
$\mathrm{BDI}(2,1;3,1)\times \mathrm{S}^2$ with the invariant subalgebra $\mathrm{SO}(2,1)\times \mathrm{SO}(3,1)\times \mathrm{U}(1)$ with 14-dimensional space-time.

To summarize, we find there are two models which are self-dual only at the classical level, these are the $\mathrm{AdS}_2\times \mathrm{S}^4$ and $\mathrm{AdS}_4\times \mathrm{S}^2$, these two model were not considered before in the context of T-self-duality and are not self-dual at the quantum level.

\begin{table}[ht]
\caption{AdS$_2$ Semi-symmetric spaces based on type-II SCA's.}
\centering
\scriptsize
\begin{tabular}{c | | c c c  c}
\hline\hline
  & OSP$(2N|2)$, $N>1$ &  D$(2,1;\a)$ &  OSP$(4^*|2N)$ & OSP$(4^*|4N)$\\ [0.5ex]
\hline \hline
$B_0$ &  U$(N|1)$ &   U$(2|1)$ &   U$(2|N)$ &  U$(2|2N)$ \\
\hline
$\begin{array}{c}
  \mathrm{Invariant }\\
  \mathrm{sub-alg}.
\end{array}$
  &  SO(N)$^2\times $U$(1)$&  U$(1)^3$ &  U$(1)^2\times$U$(N)$ &  U$(2)\times$USp$^2(2N)$ \\
   \hline
$\begin{array}{c}
  \mathrm{Bosonic-}\\
  \mathrm{subspace}
\end{array}$
&  AdS$_2\times $BDI$(N;N)$& AdS$_2\times $S$^2\times $S$^2$ &   AdS$_2\times $S$^{2}\times $CI$(N)$ &  AdS$_2\times $CII$(N;N)$ \\
               \hline
$\begin{array}{c}
  \mathrm{Rank~of}\\
  \k-\mathrm{symm}.
\end{array}$
& 0 &  0 &   0 & 8 for $N=1$ \\
  \hline
sJacobian
& $\varnothing$ &  $\varnothing$ &   $\varnothing$ &  $\neq 1$ for $N=1$\\
  \hline
Self-dual
& No &  No &  No &  $\begin{array}{c}\mathrm{Yes-C~for}~$N=1$,\\ \mathrm{else~No}\end{array}$ \\   [1ex]
\hline
\end{tabular}
\begin{quote}
Note that dim$(\mathrm{BDI}(p,q))=pq$, dim$(\mathrm{CI}(N))=N(N+1)$ and dim$(\mathrm{CII}(N;N))=4N^2$, for $N=1$ we have CII$(1;1)\simeq$ S$^4$.
\end{quote}
\label{table:Type_II_QPAdS2}
\end{table}
\normalsize

\begin{table}[ht]
\caption{AdS$_{n>3}$ Semi-symmetric spaces based on type-II SCA's.}
\centering
\scriptsize
\begin{tabular}{c | | c c c c c}
\hline\hline
  &  OSP$(4N|4)$ &  F$(4;2)$ & OSP$(8^*|2N)$ \\ [0.5ex]
\hline \hline
$B_0$ &  U$(2N|2)$ &  U$(1)\times $OSP$(2|4)$ & U$(4|N)$\\
\hline
$\begin{array}{c}
  \mathrm{Invariant }\\
  \mathrm{sub-alg}.
\end{array}$
  & U$(2N)\times $SP$(2)^2$ &  $\varnothing$ & $\varnothing$\\
   \hline
$\begin{array}{c}
  \mathrm{Bosonic-}\\
  \mathrm{subspace}
\end{array}$
&  AdS$_4\times$DIII$(2N)$ &  $\varnothing$ & $\varnothing$\\
               \hline
$\begin{array}{c}
  \mathrm{Rank~of}\\
  \k-\mathrm{symm}.
\end{array}$
&  8 for $N=1$ &  $\varnothing$ & $\varnothing$\\
  \hline
sJacobian
& $\neq 1$ for $N=1$ &  $\varnothing$ & $\varnothing$ \\
  \hline
Self-dual
& $\begin{array}{c}\mathrm{Yes-C~for}~$N=1$,\\ \mathrm{else~No}\end{array}$ &  No & No \\   [1ex]
\hline
\end{tabular}
\begin{quote}
Note that dim$(\mathrm{DIII}(N))=N(N-1)$, for $N=2$ we have DIII$(2)=$ S$^2$.
\end{quote}
\label{table:Type_II_QPAdSn}
\end{table}
\normalsize
\begin{table}[ht]
\caption{Non-AdS Semi-symmetric spaces based on type-I SCA's.}
\centering
\scriptsize
\begin{tabular}{c | | c c c c}
\hline\hline
  & OSP$(2N|4)$, $N>1$ &    OSP$(8^*|2N)$ & OSP$(8^*|4N)$ \\ [0.5ex]
\hline \hline
$B_0$ & U$(N|2)$ &   U$(4|N)$ & U$(4|2N)$ \\
\hline
$\begin{array}{c}
  \mathrm{Invariant }\\
  \mathrm{sub-alg}.
\end{array}$
  & $\mathrm{U}(2)\times \mathrm{SO}(N)^2$ &  $\mathrm{SO}(3,1)\times \mathrm{SO}(3,1)\times \mathrm{U}(N)$ & $\mathrm{U}(2,2)\times \mathrm{USp}(2N)^2$\\
   \hline
$\begin{array}{c}
  \mathrm{Bosonic-}\\
  \mathrm{subspace}
\end{array}$
& $\mathrm{CI}(4)\times \mathrm{BDI}(N;N)$ &    $\mathrm{BDI}(3,1;3,1)\times\mathrm{CI}(N)$ & $\mathrm{DIII}(2,2)\times \mathrm{CII}(N;N)$ \\
               \hline
$\begin{array}{c}
  \mathrm{Rank~of}\\
  \k-\mathrm{symm}.
\end{array}$
& 0 &   0 & 0\\
  \hline
Self-dual
& No &   No & No\\   [1ex]
\hline
\end{tabular}
\label{table:Type_II_QPnonAdS}
\end{table}
\normalsize

\subsection{AdS$_3\times \mathcal{M}$ semi-symmetric spaces}
Semi-symmetric spaces with AdS$_3$ subspace are generated by supergroups which are not simple, by taking a product of two supergroups with a bosonic subgroup of SU$(1,1)$, so we have SU$(1,1)\times $ SU$(1,1)\simeq $ SO$(2,2)$ as a subgroup. Generally we always have a $\mathbb{Z}_4$ automorphism for these products by taking the coset $\frac{G\times G}{G_{\mathrm{bosonic}}}$ \cite{Babichenko:2009dk,Zarembo:2010sg}. The cases of $\frac{\mathrm{PSU}(1,1|2)^2}{\mathrm{SU}(1,1)\times \mathrm{SU}(2)}\simeq \mathrm{AdS}_3\times \mathrm{S}^3$ and $\frac{\mathrm{D}(2,1;\a)^2}{\mathrm{SU}(1,1)\times \mathrm{SU}(2)^2}\simeq \mathrm{AdS}_3\times \mathrm{S}^3 \times \mathrm{S}^3$ are discussed in \cite{Berkovits:1999zq} and \cite{Babichenko:2009dk} respectively. Let us discuss the self-duality of these models for semi-symmetric spaces with irreducible sub-symmetric spaces. The candidates are the type I supergroups SU$(1,1|N)$ and PSU$(1,1|2)$, and the type II supergroups OSP$(N|2)$, D$(2,1;\a)$, F$(4;0)$ and G$(3;p)$.

The quotient satisfying (\ref{eq:Z4Z3condition}) with dimension $\leq 10$ are given in table \ref{table:AdS_3}. We find one self-dual model at the quantum level, $\mathrm{AdS}_3\times \mathrm{S}^3$, which was proven to be self-dual in \cite{Adam:2009kt}, and another new model self-dual only at the classical level, $\mathrm{AdS}_3\times \mathrm{S}^1$.

\begin{table}[ht]
\caption{AdS$_3$ Semi-symmetric spaces.}
\centering
\scriptsize
\begin{tabular}{c | | c c c }
\hline\hline
Background & $\mathrm{AdS}_3\times \mathrm{S}^1$ & $\mathrm{AdS}_3\times \mathrm{S}^3$ & $\mathrm{AdS}_3\times \mathrm{S}^3\times \mathrm{S}^3$\\ [0.5ex]\hline \hline
Quotient & $\frac{\mathrm{SU}(1,1|1)^2}{\mathrm{SU}(1,1)\times \mathrm{U}(1)}$ & $\frac{\mathrm{PSU}(1,1|1)^2}{\mathrm{SU}(1,1)\times \mathrm{SU}(2)}$ & $\frac{\mathrm{D}(2,1,\a)^2}{\mathrm{SU}(1,1)\times \mathrm{SO}(4)}$\\ \hline
kappa-rank & 4 & 8 & 0 \\ \hline
sJacobian & $\neq 1$ & 1 & $\varnothing$ \\ \hline
Type & I & I & II \\ \hline
Self-dual & Yes-C & Yes-Q & No \\ \hline
\end{tabular}
\begin{quote}
The osp$(4|2)$ is a special case of the $\mathrm{D}(2,1;\a)$ with $\a=1$.
\end{quote}
\label{table:AdS_3}
\end{table}
\normalsize

\subsection{AdS$_2$ semi-symmetric space}
The case where the full bosonic space AdS$_2$ is somewhat degenerate. Usually one can gauge away all fermionic degrees of freedom using kappa-symmetry, and so it comes down to T-dualizing only along one bosonic coordinate. So classically the sigma-model is self-dual, but as explained above, the dilaton will shift. For example one can construct these models using $\mathrm{OSP}(2|2)/(\mathrm{SO}(2)\times \mathrm{U}(1))$ \cite{verlinde-2004} or $\mathrm{OSP}(4|2)/(\mathrm{SO}(4)\times \mathrm{U}(1))$ \cite{Adam:2007ws}.

\subsection{When does an AdS semi-symmetric space satisfies $\O(U)=-U$?}
As one can see from the tables above, we cannot find an automorphism generating AdS subspace which also satisfies (\ref{eq:Z4Z3condition}), for all SCA's. We can see how the problem comes about when we look at the anti-commutation relations of the odd generators, and how they close on the Lorentz subalgebra, $M_{ab}$, plus noting the relations (\ref{eq:Z4DIA}) for the $\mathbb{Z}_4$ automorphism. In both cases (type-I and type-II) we have $\{Q_\a^l,S_\b^k\}\sim \d^{kl}M_{\a\b}+...$ where $\a,\b$ are the spinor indices in $d$-dimensions and $k,l$ are R-symmetry indices. When $d<6$ there are two cases where $M_{\a\b}= M_{\b\a}$ or $M_{\a\b}=-M_{\b\a}$. When $M_{\a\b}$ is antisymmetric we can just take $\O(Q_\a^l)=i S_\a^l$ and $\O(S_\a^l)=i Q_\a^l$ which gives the desired invariant bosonic subalgebra to induce AdS$_{d+1}$ space times some internal space. This is the case for $d=2$.
When $M_{\a\b}$ is symmetric we have to take $\O(Q_\a^l)=i C^l{}_k S_\a^k$ and $\O(S_\a^l)=-i C^l{}_k Q_\a^k$ where $C$ is antisymmetric full rank matrix in order to get an AdS space. This is the case for $d=3,4,5$\footnote{For $d=3$ we have $M_{\a\b}=M_{\b\a}$, $\a,\b=1,2$, spin$(3)\simeq$SU$(2)$, for $d=4$ we have self and antiself-dual generators, $M_{\a\b}=M_{\b\a}$ and $M_{\dot\a\dot\b}=M_{\dot\b\dot\a}$, $\a,\b,\dot\a,\dot\b=1,2$, spin$(4)\simeq$SU$(2)\times $SU$(2)$ and for $d=5$ we have $M_{ a b}=M_{ b a}$, $a, b=1,2,3,4$, spin$(5)\simeq$SP$(4)$. For $d=6$ we have $M_{ a b}$ traceless, $a, b=1,2,3,4$, spin$(6)\simeq$SU$(4)$.}.
Such a matrix exists only for even dimension of the R-symmetry representation of the odd generators. For $d=6$ the Lorentz generators are combinations of symmetric and anti-symmetric parts, so it is not possible to find automorphism satisfying (\ref{eq:Z4Z3condition}) which will give AdS subspace.
The details for all SCA's are found in appendix \ref{ap:Superalgebras}.

\section{Discussion}\label{sec:disscasion}
In the paper we analyzed properties of GSSM's on semi-symmetric spaces under T-duality.
For SCA's with $\mathbb{Z}$-gradation under $U$, with gradings $\pm 1,0$ only, we found three algebraic conditions that guarantee T-self-duality of the sigma-models. These are:
\begin{enumerate}
  \item $\O(U)=-U$, where $\O$ is the $\mathbb{Z}_4$ automorphism map.
  \item Rank($\k$-symmetry) $\geq$ dim$(\mathfrak{g}_{\bar 1})$/4.
  \item The SCA's Killing-form vanishes.
\end{enumerate}
We found that only three backgrounds are consistent with all three conditions. These are the $\mathrm{AdS}_n\times \mathrm{S}^n$ for $n=2,3,5$, which were found previously to be T-self-dual \cite{Berkovits:2008ic}\cite{Beisert:2008iq}\cite{Adam:2009kt}. All of these backgrounds are constructed from the type I SCA's $\mathrm{PSU}(N,N|2N)$ with $N=1,2$.

The last condition is necessary for quantum consistency of the transformation; it implies the unity of the super-Jacobian of the transformation.
We found that there are backgrounds that satisfy the first two conditions, namely they are self-dual at the level of the classical action, but their dilaton transforms non-trivially under the transformation. These are the $\mathrm{AdS}_n\times \mathrm{S}^1$ for $n=2,3,5$ which are constructed from the type I SCA's $\mathrm{SU}(N,N|N)$ with $N=1,2$ (the $N=2$ case was studied in \cite{Hao:2009hw}), and the $\mathrm{AdS}_4\times \mathrm{S}^2$ and $\mathrm{AdS}_2\times \mathrm{S}^4$ which are constructed from the type II SCA $\mathrm{OSP}(4|4)$ (and its real-forms).
All other backgrounds satisfying the first condition have rank zero kappa-symmetry.

The last condition, for quantum consistency of the transformation, also implies the vanishing of the beta-function at one-loop \cite{Kagan:2005wt}.
This might mean that having self-duality only at classical level is related to lack of worldsheet conformal invariance. If true,
adding D-branes degrees of freedom \cite{Klebanov:2004ya} may be a way to fix it.

The classification of backgrounds satisfying the first condition follows from a relation between the spinor representation of the SCA and the R-symmetry representation of the odd generators.
Besides the PSU SCA's, among the backgrounds constructed by SCA's with zero Killing-form, generating AdS spaces, (namely, $\mathrm{D}(2,1;\a)$, $\mathrm{OSP}(6|4)$ and their direct products\footnote{We include $\mathrm{OSP}(4|2)$ as a special case of $\mathrm{D}(2,1;\a)$.}) only the $\mathrm{D}(2,1;\a)$ and $\mathrm{D}(2,1;\a)^2$ SCA's admit a semi-symmetric space satisfying the first condition ($\mathrm{AdS}_2\times \mathrm{S}^2\times \mathrm{S}^2$ and $\mathrm{AdS}_3\times \mathrm{S}^3\times \mathrm{S}^3$ respectively), but these background's kappa-symmetry rank vanishes \cite{Babichenko:2009dk}\cite{Zarembo:2010sg}.
The semi-symmetric space constructed from $\mathrm{OSP}(6|4)$ that satisfies the first condition is not an AdS space, where the $\mathrm{AdS}_4\times \mathbb{CP}^3$ backgrounds doesn't satisfy it. There is also one semi-symmetric space constructed from $\mathrm{OSP}(4|2)$ which doesn't satisfy the condition which is $\mathrm{AdS}_2\times \mathrm{S}^3$.

We found there are several new families of coordinate directions (or equivalently abelian subalgebras), along which the self-dual sigma-models  also admit self-duality under T-duality summarized in figure \ref{fig:typeIdecompositions},
one of them includes only fermionic directions. These families are different for type I and type II SCA's.
The new families should generate a dual-SCA similar to the well known one \cite{Drummond:2008vq} which is associated with T-dualizing along the $P$ and $Q$ directions, but the heuristic transformation in the Yangian charges space given in \cite{Beisert:2009cs} should be with respect to the $\mathbb{Z}$-gradation charges of the generators.

For the scattering amplitudes on the gauge theory side, one associates dual variables, such that the scattering amplitudes in terms of theses variables shows manifest invariance under the dual-superconformal symmetry \cite{Drummond:2008vq}. The self-duality along different directions introduces other dual-superconformal symmetries (which of-course related to the same Yangian), so in principle we can construct dual variables in analogy to the one constructed to for T-duality along span$\{P,Q\}$ \cite{Drummond:2008vq} (see the first line in table \ref{table:DualVariables}). We give these dual-variables in table \ref{table:DualVariables}. The problem is that in terms of the on-shell variables, the dual variables include derivatives, so in order to construct them we fourier transform them. After fourier transforming the scattering amplitude, the dual variables do not appear in delta-functions as in case of T-duality along span$\{P,Q\}$, and the dual-symmetry is not manifest.

\begin{table}[ht]\label{table:DualVariables}
\caption{Dual variables.}
\centering
\scriptsize
\begin{tabular}{|c | | c |}
\hline\hline
$\begin{array}{c}
   {} \\
   {}
 \end{array}U$ & Dual variables\\ [0.5ex]
\hline \hline
$\begin{array}{c}
   {} \\
   {}
 \end{array}D+B$ & $(x_i-x_{i+1})_{\a\dot\a}=\l_{i\a}\tilde\l_{i\dot\a},\quad(\t_i-\t_{i+1})_{\a}^A=\l_{i\a}\eta^A_i$\\
\hline
$\begin{array}{c}
   {} \\
   {}
 \end{array}
2B$ & $(\t_i-\t_{i+1})_{\a}^A=\l_{i\a}\eta^A_i,\quad
(\xi_i-\xi_{i+1})_{\dot\a}^A=\m_{i\dot\a}\eta^A_i$\\
\hline
$\begin{array}{c}
   {} \\
   {}
 \end{array}D+\check R$ & $(x_i-x_{i+1})_{\a\dot\a}=\l_{i\a}\tilde\l_{i\dot\a},\quad
(\t_i-\t_{i+1})_{\a}^{A}=\l_{i\a}\eta^{A}_i,\quad
(\hat\t_i-\hat\t_{i+1})_{\dot\a A'}=\tilde\l_{i\dot\a}\psi_{i A'},\quad (r_i-r_{i+1})^A_{A'}=\eta^A_i \psi_{i A'}$\\
\hline
$\begin{array}{c}
   {} \\
   {}
 \end{array}B+\check R$ & $(\t_i-\t_{i+1})_{\a}^{A}=\l_{i\a}\eta^{A}_i,\quad
(\xi_i-\xi_{i+1})_{\dot\a}^A=\m_{i\dot\a}\eta^A_i,\quad
(r_i-r_{i+1})^A_{A'}=\eta^A_i \psi_{i A'}$\\
\hline
\end{tabular}
\begin{quote}
The on-shell variables of the $\mathcal{N}=4$ SYM scattering amplitudes are $\l,\tilde \l$ and $\eta$. $\m$ and $\psi$ are the fourier transforms of $\tilde \l$ and $\eta$ respectively.
\end{quote}
\label{table:DualVariables}
\end{table}
\normalsize

It is interesting to find if there are objects on the gauge theory side that can be related based on the other dual-superconformal algebras (similar to the scattering amplitudes/Wilson loops duality).

We showed that the flat-connection transformation under T-duality is a parameter dependent automorphism, which is the $\mathbb{Z}_4$ automorphism $\O$ followed by conjugation with $f(z)^U$ where $U$ is the generator inducing the $\mathbb{Z}$-gradation and $f$ is always the same function depending on the spectral parameter $z$.

For the case of the $\mathrm{AdS}_4\times \mathbb{CP}^3$ background, where the first condition is not satisfied, namely $\O(U)\neq - U$, there is evidence on the gauge theory side that the background may have similar properties to the $\mathrm{AdS}_5\times \mathrm{S}^5$ background which were interpreted as a consequence of T-self-duality of the background, \cite{Bargheer:2010hn}\cite{Huang:2010qy}\cite{Lee:2010du}\cite{Gang:2010gy}.
On the other hand, there is also other evidence that this background is not self-dual \cite{Adam:2009kt}\cite{Adam:2010hh}\cite{Grassi:2009yj}\cite{Bakhmatov:2010fp}.

The condition $\O(U)=-U$ seems also to be related to the Pohlmeyer-reduction of the $\mathrm{AdS}_n\times \mathrm{S}^n$ sigma-models introduced in \cite{grigoriev-2008-800}. A key property of the SCA used in the procedure of \cite{grigoriev-2008-800} was to further decompose the SCA (on top of the $\mathbb{Z}_4$ decomposition), such that a generator $T\in \mathcal{H}_2$ forms the projection $\mathcal{P}(\cdot)=[T,[T,\cdot]]$ of a $\mathbb{Z}_2$-decomposition. Since our U(1) gives charges $\pm 1$ and $0$ to all generators, taking $T=U$ the projection is $\mathcal{P}(L_a)=|a|L_a$ with $a=\pm 1,0$, so the sets $A_1\oplus A_{-1}$ have grading 1 and $B_0$ grading 0.
The condition $T\in \mathcal{H}_2$ was essential in the reduction procedure where elements in $\mathcal{H}_2$ were gauge fixed to $T$.
So actually $\O(U)=-U$ is not enough, but we also need the $\mathbb{Z}$-gradation automorphism to be inner, which is possible for all SCA's which were found to be self-dual\footnote{That is, for the PSU SCA's we can take $U=D+\check R$.}.
In cases where $T\in \mathcal{H}_0$ one might expect Pohlmeyer-reduction procedure to fail, e.g for the $\mathrm{AdS}_4\times \mathbb{CP}^3$ background.
The Pohlmeyer-reduction also relies heavily on the possibility to use kappa-symmetry as does the T-duality procedure.

\section*{Acknowledgements}

We would like to thank Ido Adam for valuable discussions and comments on the manuscript.
The work is supported in part by the Israeli
Science Foundation center of excellence, by the Deutsch-Israelische
Projektkooperation (DIP), by the US-Israel Binational Science
Foundation (BSF), and by the German-Israeli Foundation (GIF).

\appendix

\section{Notations}\label{sec:Notations}
In this section we summarize our notations used throughout the main text.

We denote the superconformal algebras (SCA's) by $\mathfrak{g}$, with the $\mathbb{Z}_2$ decomposition $\mathfrak{g}=\mathfrak{g}_{\bar 0}\oplus \mathfrak{g}_{\bar 1}$ to its even and odd parts respectively.
$\mathfrak{g}_I$, $\mathfrak{g}_{II}$ will denote type-I and Type-II SCA's respectively.

$\O$ is the $\mathbb{Z}_4$ automorphism map, which decomposes the SCA as $\mathfrak{g}=\bigoplus_{i=0}^3\mathcal{H}_i$, where $i$ denotes the grading.

The $\mathbb{Z}$-gradation decomposition with gradings $\pm1,0$ only is induced by $U$, with the charge given as the eigenvalue of $\mathrm{ad}_U$. We denote the decomposition as follows, $\mathfrak{g}=A_{-1}\oplus B_0\oplus A_1$, where the subscript indicating the grading.
The $\mathbb{Z}$-gradation also induces the map $\s_\l(L_a)=\l^U L_a \l^{-U}=\l^a L_a$, where $[U,L_a]=a L_a$, $L_a\in \mathfrak{g}$, and $a=\pm 1,0$.

We use the left-invariant-one-form, $j=g^{-1}d g$, which in the (2d-worldsheet) conformal basis is given by $j=g^{-1}\p g$, $\bar j=g^{-1}\bar \p g$. We use two decompositions of $j$, one according to the $\mathbb{Z}$-gradation, $j\equiv J+\mathrm{j}$ where $J\in A_1$ and $\mathrm{j}\in B_0$, and the other according to the $\mathbb{Z}_2$-grading, $j=j_B+j_F$ where $j_B\in \mathfrak{g}_{\bar 0}$ and $j_F\in \mathfrak{g}_{\bar 1}$.

Our indices conventions are
\begin{itemize}
  \item $I,J,K,...$ - Indices of generators in $A_1$.
  \item $\a,\b,\g,...$ - Indices of generators in $B_0$ diagonal with respect to $\O$.
  \item $A,B,C,...$ - Indices of any generator in the SCA.
\end{itemize}

We use the bilinear-form $\eta_{AB}=\Str(L_A \O(L_B))$ which is symmetric for both even and odd generators.

\section{Superalgebras}\label{ap:Superalgebras}
In this section we find the $\mathbb{Z}_4$ automorphisms satisfying (\ref{eq:Z4Z3condition}) for all SCA's with the gradations discussed in section \ref{sec:Properties of Superconformal Algebras}.
For simplicity we will not quote the entire commutation relations of the SCA's. Instead we will give only the anti-commutation relations that will suffice to constrain the automorphism. For type I and type II SCA's we assume the $\mathbb{Z}_4$ automorphism transformations (\ref{eq:Z4DIA}).
Since theses automorphisms interchanges $P$ with $K$, we must require $\O(D)=-D$.
We also note that in order to get an AdS subspace we need $\O(M_{ab})=M_{ab}$ for the Lorentz rotations.
Throughout the subsections we use the matrices $C$ and $F$, defined such that $C_{ij}C^{jk}=\d_i^k$ and $F_a{}^b F_b{}^c=\d_a^c$ where $l,k$ are R-symmetry indices and $a,b$ are spinor indices.
We assume $C_{ij}=(-)^{s_c}C_{ij}$ and $F_{ab}=(-)^{s_f}F_{ba}$ where $s_c,s_f=0$ or $1$.
We raise and lower the spinor indices using the charge conjugation matrix $\e$, $\psi^a=\e^{ a b}\psi_ b$ and $\psi_a=\e_{a b}\psi^ b$ and $\e_{ a b}\e^{ b c}=-\d_a^c$. We use the standard semi-symmetric spaces notations \cite{Helgason:2001} whenever the space is not a sphere or AdS, with a superscript/subscript indicating the dimensionality.

\subsection{OSP$(2N|2)$}
The relevant commutation relations are
\be
\{Q_l,S^k\}=\d^k_l D+\l_l{}^k,\quad
\{Q_l,\hat S_k\}=R_{lk},\quad
\{\hat Q^l,S^k\}=\hat R^{lk}.
\ee
where $l=1,...,N$ is the R-symmetry index, $\l_l{}^k$ form U$(N)$ subalgebra of SO$(2N)$, and $R_{lk}=-R_{kl}$, $\hat R^{lk}=-\hat R^{kl}$.
The automorphism transformation is
\be
\O(Q_{l})=iC_{lk}S^{k},\quad
\O(S^{l})=iC^{lk}Q_{k}.
\ee
Because $\O(D)=-D$, we must have $C_{lk}= C_{kl}$. This implies the transformation
\be
\O(C^{pl}R_{lk}\pm C_{kl}\hat R^{lp})=\mp(C^{pl}R_{lk}\pm C_{kl}\hat R^{lp}),\quad
\ee
$$
\O(C^{[pl}\l_{l}{}^{k]})=C^{[pl}\l_{l}{}^{k]},\quad
\O(C^{(pl}\l_{l}{}^{k)})=-C^{(pl}\l_{l}{}^{k)}.
$$
So the semi-symmetric space induced by the automorphism is
\be
\mathrm{AdS}_2\times \mathrm{BDI}(N;N)^{N^2}\simeq \frac{\mathrm{OSP}(2N|2)}{\mathrm{SO}(N)^2\times \mathrm{U}(1)},
\ee
which for $N=1$ is
\be
\mathrm{AdS}_2\times \mathrm{S}^{1}\simeq \frac{\mathrm{OSP}(2|2)}{\mathrm{U}(1)}.
\ee

\subsection{OSP$(2N|4)$}
The relevant commutation relations are
\be
\{Q_{l\a},S^k_\b\}=\d^k_l (\e_{\a\b}D+M_{\a\b})+\e_{\a\b}\l_l{}^k,\quad
\{Q_{l\a},\hat S_{k\b}\}=\e_{\a\b}R_{lk},\quad
\{\hat Q^l_\a,S^k_\b\}=\e_{\a\b}\hat R^{lk}.
\ee
where $l=1,...,N$ is the R-symmetry index, $\l_l{}^k$ form U$(N)$ subalgebra of SO$(2N)$, and $R_{lk}=-R_{kl}$, $\hat R^{lk}=-\hat R^{kl}$. $\a=1,2$ is the spinor index in the representation spin$(3)\simeq$ SU$(2)$ and $M_{\a\b}=M_{\b\a}$.
The automorphism transformation is
\be
\O(Q_{l\a})=iC_{lk}F_\a{}^\b S^{k}_\b,\quad
\O(S^{l}_\a)=iC^{lk}F_\a{}^\b Q_{k\b}.
\ee
We get $\O(D)=(-)^{1+s_c+s_f}D$, so we must have $C_{lk}=(-)^s C_{kl}$ and $F_{\a\b}=(-)^s F_{\b\a}$.
Only for $s=1$ we get AdS subspace (i.e, $F=\e$ and $\O(M_{\a\b})=M_{\a\b}$). If we take $s=1$ we must work with even N. The transformations of the R-symmetry are,
\be
\O(C^{pl}R_{lk}\pm C_{kl}\hat R^{lp})=\pm(C^{pl}R_{lk}\pm C_{kl}\hat R^{lp}),\quad
\ee
$$
\O(C^{[pl}\l_{l}{}^{k]})=-C^{[pl}\l_{l}{}^{k]},\quad
\O(C^{(pl}\l_{l}{}^{k)})=C^{(pl}\l_{l}{}^{k)}.
$$
So the semi-symmetric space induced by the automorphism is
\be
\mathrm{AdS}_4\times \mathrm{DIII}(2N)^{2N(2N-1)}\simeq \frac{\mathrm{OSP}(4N|4)}{\mathrm{U}(2N)\times \mathrm{SP}(2)^2}.
\ee
If we take $s=0$ we have
\be
\O(C^{pl}R_{lk}\pm C_{kl}\hat R^{lp})=\mp(C^{pl}R_{lk}\pm C_{kl}\hat R^{lp}),\quad
\ee
$$
\O(C^{[pl}\l_{l}{}^{k]})=C^{[pl}\l_{l}{}^{k]},\quad
\O(C^{(pl}\l_{l}{}^{k)})=-C^{(pl}\l_{l}{}^{k)},\quad
$$
$$
\O(M_{\a\b})=-F_\a{}^\g F_\b{}^\d M_{\d\g}.
$$
and we get the non-AdS background semi-symmetric space induced by the automorphism
\be
\mathrm{CI}(4)_6\times \mathrm{BDI}(N;N)^{N^2}\simeq \frac{\mathrm{OSP}(2N|4)}{\mathrm{SO}(N)\times \mathrm{SO}(N)\times \mathrm{U}(2)}.
\ee

\subsection{F$(4;2)$}
The relevant commutation relations are
\be
\{Q_{\a}, S_\b\}=\e_{\a\b}D+M_{\a\b}+\e_{\a\b}\l,\quad
\{Q_{\a},\hat S_{\b}\}=\e_{\a\b} R,\quad
\{\hat Q_\a,S_\b\}=\e_{\a\b} \hat R.
\ee
$\a=1,2,3,4$ is the spinor index in the representation spin$(5)\simeq$ SP$(4)$, $\e^T=-\e$ and $M_{\a\b}=M_{\b\a}$.
The automorphism transformation is
\be
\O(Q_{\a})=iF_\a{}^\b S_\b,\quad
\O(S_\a)=iF_\a{}^\b Q_{k\b}.
\ee
We get $\O(D)=(-)^{1+s_f}D$, so we must have $F_{\a\b}=F_{\b\a}$.
But then we have $\O(F_{[\g}{}^\a M_{\a\b]})=F_{[\g}{}^\a M_{\a\b]}$, so the invariant subalgebra contains just part of the Lorentz subalgebra so$(1,4)$, namely the subalgebra $F_{[\g}{}^\a M_{\a\b]}\simeq\mathrm{u}(1,1) \in\mathcal{H}_0$.
So we get the non-AdS semi-symmetric space
\be
\mathrm{BDI}(3,1;2,1)_{12}\times \mathrm{S}^2 \simeq \frac{\mathrm{F}(4;2)}{\mathrm{SO}(3,1)\times \mathrm{SO}(2,1)\times \mathrm{U}(1)}.
\ee

\subsection{F$(4;0)$}
The relevant commutation relations are
\be
\{Q_{\a}, S_\b\}=\e_{\a\b}D+\l_{\a\b}+\e_{\a\b}\l,\quad
\{Q_{\a},\hat S_{\b}\}=R_{\a\b},\quad
\{\hat Q_\a, S_\b\}=\hat R_{\a\b}.
\ee
$\a,\b=1,2,3,4$ is the spinor index in the representation of the R-symmetry spin$(5)\simeq$ SP$(4)$, $\l_{\a\b}=(\e\G^a\G^b)_{\a\b}N_{ab}=\l_{\b\a}$ is ten-dimensional and $R_{\a\b}=(\e\G^a)_{\a\b} A_a=-R_{\b\a}$ and $\hat R_{\a\b}=(\e\G^a)_{\a\b} B_a=-\hat R_{\b\a}$ are five-dimensional (the $\G$'s are $4\times 4$ gamma-matrices in five dimensions, $a=1,...,5$, we have the constraint $\Tr(\e R)=\Tr(\e \hat R)=0$).
The automorphism transformation is
\be
\O(Q_{\a})=iF_\a{}^\b  S_\b,\quad
\O( S_\a)=iF_\a{}^\b Q_{\b}.
\ee
We get $\O(D)=(-)^{1+s_f}D$, so we must have $F_{\a\b}=F_{\b\a}$.
Then we have $\O(F_{[\g}{}^\a \l_{\a\b]})=F_{[\g}{}^\a \l_{\a\b]}$,
so the invariant subalgebra contains just part of the subalgebra $\l_{\a\b}\simeq$ so$(5)$, namely the subalgebra $F_{[\g}{}^\a \l_{\a\b]}\simeq\mathrm{u}(2) \in\mathcal{H}_0$. We also have $\O(\l)=-\l$ and $\O(F_\g{}^\a R_{\a\b})=-F_\b{}^\a \hat R_{\a\g}$.
So the semi-symmetric space is
\be
\mathrm{AdS}_{2}\times \mathrm{BDI}(3;4)^{12} \simeq \frac{\mathrm{F}(4;0)}{\mathrm{SO}(3)\times \mathrm{SO}(4)\times \mathrm{U}(1)}.
\ee

\subsection{SU$(1,1|N)$, $N\neq 2$}
The relevant commutation relations are
\be
\{Q_l, S^k\}=\d^k_l (D+A)+\l_l{}^k,\quad
\{Q_l,S_k\}=0,\quad
\{\hat Q^l,\hat S^k\}=0,
\ee
where $l=1,...,N$ is the R-symmetry index, $(\l_l{}^k)^\dag=-\l_k{}^l$ are the SU$(N)$ generators, and $A$ is the U$(1)$.
The automorphism transformation is
\be
\O(Q_{l})=iC_{lk} S^{k},\quad
\O(S^{l})=iC^{lk}Q_{k}.
\ee
In order to have $\O(D)=-D$, we must take $C_{lk}= C_{kl}$. This implies the transformation
\be
\O(C^{[pl}\l_{l}{}^{k]})=C^{[pl}\l_{l}{}^{k]},\quad
\O(C^{(pl}\l_{l}{}^{k)})=-C^{(pl}\l_{l}{}^{k)},\quad
\O(A)=-A.
\ee
So SO$(N)\simeq C^{[pl}\l_{l}{}^{k]}\in H_0$.
This implies the semi-symmetric space induced by the DIA is
\be
\mathrm{AdS}_2\times \mathrm{AI}(N)^{\frac{(N-1)(N+2)}{2}}\times \mathrm{S}^1\simeq \frac{\mathrm{SU}(1,1|N)}{\mathrm{U}(1)\times \mathrm{SO}(N)}.
\ee

\subsection{SU$(2,2|N)$, $N\neq 4$}
The relevant commutation relations are
\be
\{Q_{l\a},S^k_\b\}=\d^k_l (\e_{\a\b}(D+A)+M_{\a\b})+\e_{\a\b}\l_l{}^k,\quad
\{Q_{l\a},S_{k\b}\}=0,\quad
\{\hat Q^l_{\dot\a},\hat S^k_{\dot \b}\}=0,
\ee
where $l=1,...,N$ is the R-symmetry index, $(\l_l{}^k)^\dag=-\l_k{}^l$ are the SU$(N)$ generators, and $A$ is the U$(1)$. $\a=1,2$ and $\dot\a=1,2$ are the spinor indices in the representation spin$(4)\simeq$SU$(2)\times $SU$(2)$ and $M_{\a\b}=M_{\b\a}$ and we also have $M_{\dot\a\dot\b}=M_{\a\b}^\dag$.
The automorphism transformation is
\be
\O(Q_{l\a})=iC_{lk}F_\a{}^\b  S^{k}_\b,\quad
\O( S^{l}_\a)=iC^{lk}F_\a{}^\b Q_{k\b}.
\ee
We get $\O(D)=(-)^{1+s_c+s_f}D$, so we must have $C_{lk}=(-)^s C_{kl}$ and $F_{\a\b}=(-)^s F_{\b\a}$.
Only for $s=1$ we get AdS subspace ($F=\e$). If we take $s=1$ we must work with even N. The transformations of the R-symmetry are,
\be
\O(C^{(ml}\l_{l}{}^{k)})=C^{(ml}\l_{l}{}^{k)},\quad
\O(C^{[ml}\l_{l}{}^{k]})=-C^{[ml}\l_{l}{}^{k]},\quad
\O(A)=-A.
\ee
So USp$(N)\simeq C^{(pl}\l_{l}{}^{k)}\in H_0$,
and the semi-symmetric space induced by the transformation is
\be
\mathrm{AdS}_5\times \mathrm{AII}(N)^{(N-1)(2N+1)}\times \mathrm{S}^1\simeq \frac{\mathrm{SU}(2,2|2N)}{\mathrm{USp}(2,2)\times \mathrm{USp}(2N) }.
\ee
If we take $s=0$ we have
\be
\O(C^{(ml}\l_{l}{}^{k)})=-C^{(ml}\l_{l}{}^{k)},\quad
\O(C^{[ml}\l_{l}{}^{k]})=C^{[ml}\l_{l}{}^{k]},\quad
\O(A)=-A,
\ee
$$
\O(M_{\a\b})=-F_\a{}^\g F_\b{}^\d M_{\d\g},\quad
\O(M_{\dot \a\dot \b})=-F_{\dot \a}{}^{\dot \g} F_{\dot \b}{}^{\dot \d} M_{\dot \d\dot \g},\quad
$$
so SO$(N)\simeq C^{[pl}\l_{l}{}^{k]}\in H_0$, and we get the non-AdS background semi-symmetric space induced by the automorphism
\be
\mathrm{AI(2,2)}_6\times \mathrm{AI(N)}^{\frac{(N-1)(N+2)}{2}}\times \mathrm{S}^1\simeq\frac{\mathrm{SU}(2,2|N)}{\mathrm{SO}(2,2)\times \mathrm{SO}(N)}
\ee

\subsection{PSU$(N,N|2N)$, $N=1,2$}
The PSU superalgebras have the same structure as the SU superalgebras with the (important) modification of eliminating the S$^1$.
For PSU$(1,1|2)$ we get
\be
\mathrm{AdS}_2\times \mathrm{S}^2\simeq \frac{\mathrm{PSU}(1,1|2)}{\mathrm{U}(1)\times \mathrm{SO}(2)}.
\ee
For PSU$(2,2|4)$ we get
\be
\mathrm{AdS}_5\times \mathrm{S}^5\simeq \frac{\mathrm{PSU}(2,2|4)}{\mathrm{USp}(2,2)\times \mathrm{USp}(4) },
\ee
or
\be
\mathrm{AI(2,2)}_6\times \mathrm{AI(4)}^{9} \simeq\frac{\mathrm{PSU}(2,2|N)}{\mathrm{SO}(2,2)\times \mathrm{SO}(4)}.
\ee

In the case of $\mathrm{AdS}_2\times \mathrm{S}^2$ the rank of the $\k$-symmetry is 4 \cite{Berkovits:1999zq}.
In the case of $\mathrm{AdS}_5\times \mathrm{S}^5$ the rank of the $\k$-symmetry is 16 \cite{Metsaev:1998it}.
In the case of $\mathrm{AI(2,2)}_6\times \mathrm{AI(4)}^{9}$ the rank of the $\k$-symmetry is 0.

\subsection{D$(2,1;\z)$}
The relevant commutation relations are
\be
\{Q_\a, S_\b\}=\e_{\a\b}D+\e_{\a\b}\l+\l_{\a\b},\quad
\ee
\be
\{Q_\a,\hat S_\b\}=\e_{\a\b}R,\quad
\{\hat Q_\a, S_\b\}=\e_{\a\b}\hat R,\quad
\ee
where $\z=\cos^2(\phi)$, $\l=\cos^2{\phi}L_3$ and $\l_{\a\b}=\l_{\b\a}=-\sin^2{\phi}(\e\g^a R_a)_{\a\b}\simeq \mathrm{su}(2)$, $R=-\cos^2{\phi}L_+$ and $\hat R=\cos^2{\phi}L_-$. $\a=1,2$ is a spinor index of the R-symmetry spin$(2)$.
The automorphism transformation is
\be
\O(Q_{\a})=iF_\a{}^\b  S_\b,\quad
\O( S_\a)=iF_\a{}^\b Q_{k\b}.
\ee
We get $\O(D)=(-)^{1+s_f}D$, so we must have $F_{\a\b}=F_{\b\a}$.
Then we have $\O(F_{[\g}{}^\a \l_{\a\b]})=F_{[\g}{}^\a \l_{\a\b]}$, so U$(1)\simeq F_{[\g}{}^\a \l_{\a\b]}\in\mathcal{H}_0$. We also have $\O(\l)=-\l$ and $\O(R)=-\hat R$.
The semi-symmetric space is
\be
\mathrm{AdS}_2\times \mathrm{S}^2\times \mathrm{S}^2\simeq \frac{\mathrm{D}(2,1;\a)}{\mathrm{U}(1)\times \mathrm{U}(1)\times \mathrm{U}(1)}.
\ee

The $\k$-symmetry rank is zero \cite{Zarembo:2010sg}.

\subsection{OSP$(4^*|2N)$}
The relevant commutation relations are
\be
\{Q_{l\a}, S^k_\b\}=\e_{\a\b}\d_l^k D+\e_{\a\b}\l_l{}^k+\d_l^k \l_{\a\b},\quad
\ee
\be
\{Q_{l\a},\hat S_{k\b}\}=\e_{\a\b}R_{lk},\quad
\{\hat Q^l_\a, S^k_\b\}=\e_{\a\b}\hat R^{lk},\quad
\ee
$\l_{\a\b}=\l_{\b\a}$ form SU$(2)$ subalgebra of SO$^*(4)$, $\l_l{}^k$ forms U$(N)$ subalgebra of USp$(2N)$, $R_{lk}=R_{kl}$ and $\hat R^{lk}=\hat R^{kl}$ ($l,k=1,...,N$). $\a=1,2$ is a spinor index of the R-symmetry spin$(2)$.
The automorphism transformation is
\be
\O(Q_{l\a})=iC_{lk}F_\a{}^\b  S^{k}_\b,\quad
\O( S^{l}_\a)=iC^{lk}F_\a{}^\b Q_{k\b}.
\ee
We get $\O(D)=(-)^{1+s_c+s_f}D$, so we must have $C_{lk}=(-)^s C_{kl}$ and $F_{\a\b}=(-)^s F_{\b\a}$.
Only for $s=1$ we get AdS subspace ($F=\e$). If we take $s=1$ we must work with even N. The transformations of the R-symmetry are,
\be
\O(C^{pl}R_{lk}\pm C_{kl}\hat R^{lp})=\pm(C^{pl}R_{lk}\pm C_{kl}\hat R^{lp}),\quad
\ee
$$
\O(C^{(ml}\l_{l}{}^{k)})=C^{(kl}\l_{l}{}^{m)},\quad
\O(C^{[ml}\l_{l}{}^{k]})=-C^{[kl}\l_{l}{}^{m]},\quad
$$
$$
\O(F_{[\g}{}^\a \l_{\a\b]})=-F_{[\g}{}^\a \l_{\a\b]},\quad
\O(F_{(\g}{}^\a \l_{\a\b)})=F_{(\g}{}^\a \l_{\a\b)}.
$$
So the invariant subalgebra under the automorphism includes SP$(N)\simeq C^{(ml}\l_{l}{}^{k)}$ and SU$(2)\simeq F_{(\g}{}^\a \l_{\a\b)}$.
Thus, the semi-symmetric space induced by the automorphism is
\be
\mathrm{AdS}_2\times \mathrm{CII}(N;N)^{4N^2}\simeq \frac{\mathrm{OSP}(4^*|4N)}{\mathrm{U}(2)\times \mathrm{SP}(2N)^2},
\ee
note that CII$(1;1)\simeq$S$^4$.\\
If we take $s=0$ we get
\be
\O(C^{pl}R_{lk}\pm C_{kl}\hat R^{lp})=\mp(C^{pl}R_{lk}\pm C_{kl}\hat R^{lp}),\quad
\ee
$$
\O(C^{(ml}\l_{l}{}^{k)})=-C^{(kl}\l_{l}{}^{m)},\quad
\O(C^{[ml}\l_{l}{}^{k]})=C^{[kl}\l_{l}{}^{m]},\quad
$$
$$
\O(F_{[\g}{}^\a \l_{\a\b]})=F_{[\g}{}^\a \l_{\a\b]},\quad
\O(F_{(\g}{}^\a \l_{\a\b)})=-F_{(\g}{}^\a \l_{\a\b)}.
$$
So now the invariant subalgebra under the automorphism includes SO$(N)\simeq C^{[ml}\l_{l}{}^{k]}$ and U$(1)\simeq F_{[\g}{}^\a \l_{\a\b]}$.
the semi-symmetric space
\be
\mathrm{AdS}_2\times \mathrm{S}^2\times \mathrm{CI}(N)^{N(N+1)}\simeq \frac{\mathrm{OSP}(4^*|2N)}{\mathrm{U}(1)\times \mathrm{U}(1)\times \mathrm{U}(N)},
\ee
note that CI$(1)\simeq$ S$^2$ which is similar to the D$(2,1;\z)$ result.
\subsection{OSP$(8^*|2N)$}
The relevant commutation relations are
\be
\{Q_{l\a}, S^k_\b\}=\d_l^k (\e_{\a\b}D+M_{\a\b})+\e_{\a\b}\l_l{}^k,\quad
\ee
\be
\{Q_{l\a},\hat S_{k\b}\}=\e_{\a\b}R_{lk},\quad
\{\hat Q^l_\a, S^k_\b\}=\e_{\a\b}\hat R^{lk},\quad
\ee
$\l_l{}^k$ forms U$(N)$ subalgebra of USp$(2N)$, $R_{lk}=R_{kl}$ and $\hat R^{lk}=\hat R^{kl}$ ($l,k=1,...,N$). $\a=1,...,4$ is a spinor index of the R-symmetry spin$(6)\simeq$ SU$(4)$. $M_{\a\b}$ is the 15-dimensional SO$(1,5)\simeq$SU$^*(4)$ Lorentz subalgebra,
so $\S M_{\a\a}=0$.
Since $M_{\a\b}$ is neither symmetric nor antisymmetric we'll not be able to get an AdS space satisfying (\ref{eq:Z4Z3condition}).
The automorphism transformation is
\be
\O(Q_{l\a})=iC_{lk}F_\a{}^\b \hat S^{k}_\b,\quad
\O(\hat S^{l}_\a)=iC^{lk}F_\a{}^\b Q_{k\b}.
\ee
We get $\O(D)=(-)^{1+s_c+s_f}D$, so we must have $C_{lk}=(-)^s C_{kl}$ and $F_{\a\b}=(-)^s F_{\b\a}$.
If we take $s=1$ we must work with even N. The transformations of the R-symmetry are,
\be
\O(C^{pl}R_{lk}\pm C_{kl}\hat R^{lp})=\pm(C^{pl}R_{lk}\pm C_{kl}\hat R^{lp}),\quad
\ee
$$
\O(C^{(ml}\l_{l}{}^{k)})=C^{(ml}\l_{l}{}^{k)},\quad
\O(C^{[ml}\l_{l}{}^{k]})=-C^{[ml}\l_{l}{}^{k]},\quad
$$
$$
\O(F_{(\z}{}^\a M_{\a\b)})=F_{(\z}{}^\a M_{\a\b)},\quad
\O(F_{[\z}{}^\a M_{\a\b]})=-F_{[\z}{}^\a M_{\a\b]}.
$$
So the invariant subalgebra contains $F_{(\z}{}^\a M_{\a\b)}\simeq \mathrm{SP}(4)$ and $C^{(ml}\l_{l}{}^{k)}\simeq \mathrm{USp}(N)$.
Thus, the semi-symmetric space induced by the automorphism is
\be
\mathrm{DIII}(2,2)_{12}\times \mathrm{CII}(N;N)^{4N^2}\simeq \frac{\mathrm{OSP}(8^*|4N)}{\mathrm{U}(2,2)\times \mathrm{USp}(2N)^2},
\ee
note that CII$(1;1)\simeq$S$^4$.
If we take $s=0$ we get
\be
\O(C^{pl}R_{lk}\pm C_{kl}\hat R^{lp})=\mp(C^{pl}R_{lk}\pm C_{kl}\hat R^{lp}),\quad
\ee
$$
\O(C^{(ml}\l_{l}{}^{k)})=-C^{(ml}\l_{l}{}^{k)},\quad
\O(C^{[ml}\l_{l}{}^{k]})=C^{[ml}\l_{l}{}^{k]},\quad
$$
$$
\O(F_{(\z}{}^\a M_{\a\b)})=-F_{(\z}{}^\a M_{\a\b)},\quad
\O(F_{[\z}{}^\a M_{\a\b]})=F_{[\z}{}^\a M_{\a\b]}.
$$
So the invariant subalgebra contains $F_{[\z}{}^\a M_{\a\b]}\simeq \mathrm{SO}(2,1)\times \mathrm{SO}(3)$ and $C^{[ml}\l_{l}{}^{k]}\simeq \mathrm{SO}(N)$.
Thus, the semi-symmetric space
\be
\mathrm{BDI}(3,1;3,1)_{16}\times \mathrm{CI}(N)^{N(N+1)}\simeq \frac{\mathrm{OSP}(8^*|2N)}{\mathrm{SO}(3,1)\times \mathrm{SO}(3,1)\times \mathrm{U}(N)},
\ee
note that CI$(1)\simeq$S$^2$.

\section{Rank of kappa symmetry}\label{ap:kappa}
In this section we calculate the rank of the kappa symmetry following \cite{Zarembo:2010sg}. We shall also use the same notations for the different type of $\mathbb{Z}_4$'s. Most of the calculations are the same as in \cite{Zarembo:2010sg} where we just have to modify the dimension of the matrices.

The rank of the kappa-symmetry is $N_\k+N_{\tilde \k}$ where
\be
N_\k=\mathrm{dim}~\mathrm{ker}~\mathrm{ad}~K|_{\mathcal{H}_3}, \quad
N_{\tilde \k}=\mathrm{dim}~\mathrm{ker}~\mathrm{ad}~\bar K|_{\mathcal{H}_1},
\ee
with $K$ and $\bar K$ general elements in $\mathcal{H}_2$, which will be denoted as
\be
K(\mathrm{or}~\bar K)=\left(
  \begin{array}{cc}
    A & 0\\
    0 & B \\
  \end{array}
\right).
\ee
We'll compute
\be
\left[
\left(
  \begin{array}{cc}
    A & 0\\
    0 & B \\
  \end{array}\right),
  \left(
  \begin{array}{cc}
    0 & \Theta \\
    \Psi & 0 \\
  \end{array}
\right)
\right]=
\left(
  \begin{array}{cc}
    0 & A\Theta -\Theta B\\
    B\Psi -\Psi A & 0 \\
  \end{array}\right)
\ee
so the commutator vanish if $A\Theta=\Theta B$ and $B\Psi=\Psi A$, and the number of solution to this equation is the rank of the kappa-symmetry.
In addition we should impose the Virasoro constraint
\be
\mathrm{tr} A^2= \mathrm{tr} B^2.
\ee

We use the notation of \cite{Zarembo:2010sg} for the different types of semi-symmetric spaces. The notation type-U2 and type-U4 refers to the $\mathbb{Z}_4$ gradings of SU$(M|N)$ or PSU$(M|N)$ superalgebras, where the invariant locos
is $\mathrm{SO}(M)\times \mathrm{SO}(N)$ and $\mathrm{SP}(M)\times \mathrm{SP}(N)$ respectively, type-O1 and type-O2 refers to OSP$(M|N)$ with the invariant locus $\mathrm{SO}(2M-P)\times \mathrm{SO}(P)\times \mathrm{U}(N)$ and $\mathrm{U}(M)\times \mathrm{SP}(2N-2P)\times \mathrm{SP}(2P)$ respectively, and type-Tu and type-To to $\mathrm{(P)SU}(M|N)^2/\mathrm{SU}(M)\times \mathrm{SU}(N)$ and $\mathrm{OSP}(M|2N)^2/\mathrm{SO}(M)\times \mathrm{SP}(2N)$ respectively.

\subsection{Type-U2}
The coset space is
\be\label{eq:cosetO1}
\mathrm{AI}(N)_{\frac{(N-1)(N+2)}{2}}\times \mathrm{S}^1\times \mathrm{AI}(M)^{\frac{(M-1)(M+2)}{2}}\simeq\frac{\mathrm{SU}(M|N)}{\mathrm{SO}(M)\times \mathrm{SO}(N)}.
\ee
The relevant models in our classification are
\be
\begin{array}{l}
  \mathrm{AdS}_2\times \mathrm{S}^1\times \mathrm{AI}(N)^{\frac{(N-1)(N+2)}{2}}\simeq\frac{\mathrm{SU}(1,1|N)}{\mathrm{SO}(1,1)\times \mathrm{SO}(N)},\quad
N\neq 2, \\
\\
  \mathrm{AdS}_2\times \mathrm{S}^2\simeq\frac{\mathrm{PSU}(1,1|2)}{\mathrm{SO}(1,1)\times \mathrm{SO}(2)},\quad \\
  \\
  \mathrm{AI}(2,2)_9\times \mathrm{S}^1\times \mathrm{AI}(N)^{\frac{(N-1)(N+2)}{2}}\simeq\frac{\mathrm{SU}(2,2|N)}{\mathrm{SO}(2,2)\times \mathrm{SO}(N)},\quad
N\neq 4, \\
\\
  \mathrm{AI}(2,2)_9\times \mathrm{AI}(4)^{9}\simeq\frac{\mathrm{PSU}(2,2|4)}{\mathrm{SO}(2,2)\times \mathrm{SO}(4)}.
\end{array}
\ee

The calculation of the rank goes the same as in  \cite{Zarembo:2010sg}, where one have to solve
\be
A\Theta=\Theta B,\quad
A=A^t,\quad
B=B^t,\quad
\Tr(A)=\Tr(B)(=0~\mathrm{ ~for~ PSU's}),\quad
\ee
$$
\Tr(A^2)=\Tr(B^2).
$$
There are no solutions in general to this equation and the rank is zero, with the two exceptions:
\begin{itemize}
  \item $\mathrm{PSU}(1,1|2)$ - in which there are two solutions, so the rank is four, namely, $N_\k=N_{\tilde\k}=2$ \cite{Zarembo:2010sg}. The coset space is $\mathrm{AdS}_2\times \mathrm{S}^2$.
  \item $\mathrm{SU}(1,1|1)$ - in which there is one solution, so the rank is two, namely, $N_\k=N_{\tilde\k}=1$. The eigenvalues of $A$ and B are $\{0,\a\}$ and $\{\a\}$ respectively. The coset space is $\mathrm{AdS}_2\times \mathrm{S}^1$.
\end{itemize}
In both cases the kappa symmetry rank is half the number of fermionic d.o.f.
\subsection{Type-U4}
The coset space is
\be\label{eq:cosetO1}
\mathrm{AII}(M)_{(M-1)(2M+1)}\times \mathrm{S}^1\times \mathrm{AII}(N)^{(N-1)(2N+1)}\simeq\frac{\mathrm{SU}(M|N)}{\mathrm{SP}(M)\times \mathrm{SP}(N)}.
\ee
The relevant models in our classification,
\be
\begin{array}{l}
  \mathrm{AdS}_5\times \mathrm{S}^1\times \mathrm{AII}(N)^{(N-1)(2N+1)}\simeq\frac{\mathrm{SU}(2,2|2N)}{\mathrm{USp}(2,2)\times \mathrm{USp}(2N)},\quad
N\neq 2, \\
   \\
  \mathrm{AdS}_5\times \mathrm{S}^{5}\simeq\frac{\mathrm{PSU}(2,2|4)}{\mathrm{USp}(2,2)\times \mathrm{USp}(4)}.\quad \\
\end{array}
\ee

The calculation of the rank goes the same as in  \cite{Zarembo:2010sg}, where one have to solve
\be
A\Theta=\Theta B,\quad
A=-J A^t J,\quad
B=-\tilde J B^t\tilde J ,\quad
\Tr(A)=\Tr(B)(=0~\mathrm{ ~for~ PSU's}),\quad
\ee
$$
\Tr(A^2)=\Tr(B^2).
$$
There are no solutions in general to this equation and the rank is zero, with the two exceptions:
\begin{itemize}
  \item $\mathrm{PSU}(2,2|4)$ - in which there are eight solutions, so the rank is sixteen, namely, $N_\k=N_{\tilde\k}=8$ \cite{Zarembo:2010sg}.
  \item $\mathrm{SU}(2,2|2)$ - in which there are four solution, so the rank is eight, namely, $N_\k=N_{\tilde\k}=4$. The eigenvalues of $A$ and B are $\{0,0,\a,\a\}$ and $\{\a,\a\}$ respectively.
\end{itemize}
In both cases the kappa symmetry rank is half the number of fermionic d.o.f.

\subsection{Type-O1}
The relevant coset is
\be\label{eq:cosetO1}
\mathrm{BDI}(2M-P;P)_{P(2M-P)}\times \mathrm{CI}(N)^{N(N+1)}\simeq\frac{\mathrm{OSP}(2M|2N)}{\mathrm{SO}(2M-P)\times \mathrm{SO}(P)\times \mathrm{U}(N)}.
\ee
We have several models relevant for our classification (both type I and type II),
\be
\begin{array}{l}
\mathrm{CI}(4)_6\times \mathrm{S}^1\simeq\frac{\mathrm{OSP}(2|4)}{\mathrm{U}(2)},\quad\\
\\
\mathrm{AdS}_2\times \mathrm{BDI}(N;N)^{N^2} \simeq\frac{\mathrm{OSP}(2N|2)}{\mathrm{SO}(N)^2\times \mathrm{U}(1)},\quad\\
\\
\mathrm{AdS}_2\times \mathrm{S}^2\times \mathrm{CI}(N)^{N(N+1)}\simeq\frac{\mathrm{OSP}(4^*|2N)}{\mathrm{SO}(2)^2\times \mathrm{U}(N)},\quad\\
\\
\mathrm{CI}(4)_6\times\mathrm{BDI}(N;N)^{N^2} \simeq\frac{\mathrm{OSP}(2N|4)}{\mathrm{SO}(N)^2\times \mathrm{U}(2)},\quad\\
\\
\mathrm{BDI}(3,1;3,1)_{16}\times \mathrm{CI}(N)^{N(N+1)}\simeq\frac{\mathrm{OSP}(8^*|2N)}{\mathrm{SO}(3,1)^2\times \mathrm{U}(N)}.\quad\\
\end{array}
\ee

The calculation of the rank is very similar to the one in \cite{Zarembo:2010sg}. In all of our cases we have $P=M$ in (\ref{eq:cosetO1}), so we'll analyze the rank of these model only.
One has the $\mathbb{Z}_4$ decomposition
\be
\begin{array}{cc}
  \mathcal{H}_2: & A=\left(
                        \begin{array}{cc}
                          0 & A_{1(M\times M)} \\
                          -A^t_{1(M\times M)} & 0 \\
                        \end{array}
                      \right),\quad
                    B=\left(
                        \begin{array}{cc}
                          B_{1(N\times N)} & B_{2(N\times N)} \\
                          B_{2(N\times N)} & -B_{1(N\times N)} \\
                        \end{array}
                      \right),\quad                      B_i=B_i^t,
   \\
  \mathcal{H}_1: & \left(
                        \begin{array}{cc}
                          \T_{1(M\times N)} & -i\T_{1(M\times N)} \\
                          \T_{2(M\times N)} & i\T_{1(M\times N)} \\
                        \end{array}
                      \right),\quad \\
  \mathcal{H}_3: & \left(
                        \begin{array}{cc}
                          \T_{1(M\times N)} & i\T_{1(M\times N)} \\
                          \T_{2(M\times N)} & -i\T_{1(M\times N)} \\
                        \end{array}
                      \right).
\end{array}
\ee
The equation $A\T=\T B$ and the Virasoro constraint comes down to solving
\be
-A_1 A_1^t\T_1=\T_1 B_+ B_-, \quad B_\pm=B_1\pm i B_2,\quad
-2\Tr(A_1 A_1^t)=2\Tr(B_+ B_-).
\ee
In general there are no solutions to this equation and hence the rank of kappa-symmetry is zero. There is one exception
\begin{itemize}
  \item OSP$(2|2)$ - in this case $A_1$ and $B_{\pm}$ are numbers and there is one solution so the rank of kappa-symmetry is two, namely, $N_\k=N_{\tilde\k}=1$ (this is actually the same result as for the type-U2 SU$(1,1|1)$). In this case the kappa symmetry rank is half the number of fermionic d.o.f.
\end{itemize}

\subsection{Type-O2}
The relevant coset is
\be\label{eq:cosetO2}
\mathrm{DIII}(M)_{M(M-1)}\times \mathrm{CII}(N-P;P)^{4P(N-P)}\simeq
\frac{\mathrm{OSP}(2M|2N)}{\mathrm{U}(M)\times \mathrm{SP}(2N-2P)\times \mathrm{SP}(2P)}.
\ee
We have several models relevant for our classification,
\be
\begin{array}{l}
\mathrm{AdS}_2\times \mathrm{CII}(N;N)^{4N^2}\simeq\frac{\mathrm{OSP}(4^*|4N)}{\mathrm{U}(2)\times SP(2N)^2},\quad\\
\\
\mathrm{AdS}_4\times \mathrm{DIII}(2N)^{2N(2N-1)}\simeq\frac{\mathrm{OSP}(4N|4)}{\mathrm{U}(2N)\times SP(2)^2},\quad\\
\\
\mathrm{DIII}(2;2)_{12}\times \mathrm{CII}(N;N)^{4N^2}\simeq\frac{\mathrm{OSP}(8^*|4N)}{\mathrm{U}(4)\times SP(2N)^2}.\quad\\
\end{array}
\ee

The calculation of the rank is very similar to the one in \cite{Zarembo:2010sg}. In all of our cases we have $N$ and $M$-even and $P=N/2$ in (\ref{eq:cosetO2}), so we'll analyze the rank of these model only.
One has the $\mathbb{Z}_4$ decomposition
\be
\begin{array}{cc}
  \mathcal{H}_2: & A=\left(
                        \begin{array}{cc}
                          A_{1(M\times M)} & A_{2(M\times M)} \\
                          A_{2(M\times M)} & -A_{1(M\times M)} \\
                        \end{array}
                      \right),\quad        A_i=-A_i^t,\quad
                    B=\left(
                        \begin{array}{cc}
                          0_{(N\times N)} & B_{1(N\times N)} \\
                          J B^t_{1(N\times N)} J & 0_{(N\times N)} \\
                        \end{array}
                      \right),\quad
   \\
  \mathcal{H}_1: & \left(
                        \begin{array}{cc}
                          \T_{1(M\times N)} & \T_{2(M\times N)} \\
                          -i\T_{1(M\times N)} & i\T_{2(M\times N)} \\
                        \end{array}
                      \right),\quad \\
  \mathcal{H}_3: & \left(
                        \begin{array}{cc}
                          \T_{1(M\times N)} & \T_{2(M\times N)} \\
                          i\T_{1(M\times N)} & -i\T_{2(M\times N)} \\
                        \end{array}
                      \right).
\end{array}
\ee
The equation $A\T=\T B$ and the Virasoro constraint comes down to solving
\be
-A_\pm A_\mp^t\T_1=\T_1 B_1 J B_1^t J,\quad
\Tr(A_1^2+A_2^2)=\Tr(B_1 J B_1^t J).
\ee
In general there are no solutions to this equation and hence the rank of kappa-symmetry is zero. There is one exception
\begin{itemize}
  \item OSP$(4|4)$ - in this case the eigenvalues of $-A_\pm A_\mp^t$ and $B_1 J B_1^t J$ are $\{\a,\a\}$ and $\{\a,\a\}$, so there are four solutions, so the rank of the kappa-symmetry is eight, namely, $N_\k=N_{\tilde\k}=4$. The backgrounds in this case are $\mathrm{AdS}_4\times \mathrm{S}^2$ or $\mathrm{AdS}_2\times \mathrm{S}^4$ if we take the real form of the superalgebra. In this case the kappa-symmetry rank is half the number of fermionic d.o.f.
\end{itemize}

\subsection{Type-Tu}
The coset space is
\be\label{eq:cosetTu}
\frac{\mathrm{SU}(M|N)^2}{\mathrm{SU}(M)\times \mathrm{SU}(N)\times \mathrm{U}(1)}.
\ee
In our classification we are interested only in AdS$_3$ models, so we take $M=2$ (or more precisely $SU(1,1)$).
The relevant models in our classification are
\be
\begin{array}{l}
\mathrm{AdS}_3\times \mathrm{S}^1\times \frac{\mathrm{SU}(N)^2}{\mathrm{SU}(N)}\simeq\frac{\mathrm{SU}(1,1|N)^2}{\mathrm{SU}(1,1)\times \mathrm{U}(N)},\quad
N\neq 2,\\
\\
\mathrm{AdS}_3\times \mathrm{S}^{3}\simeq\frac{\mathrm{PSU}(1,1|2)}{\mathrm{SU}(1,1)\times \mathrm{SU}(2)}.\quad\\
\end{array}
\ee

According to \cite{Zarembo:2010sg} we have to find the number of solutions to the two equations $A\T=\T B$ and $B\Psi=\Psi A$, so we need to find out how many eigenvalues of $A=\mathrm{U}(2)$ and $B=\mathrm{U}(M)$ coincide upon the Virasoro constraint and $\Tr(A)=\Tr(B)$. In general the eigenvalues are different and we don't have kappa-symmetry.
There are two exceptions
\begin{itemize}
  \item PSU$(1,1|2)$ - in this case the eigenvalues of $A$ and $B$ are $\{\a,-\a\}$ and $\{\a,-\a\}$, so there are two solutions for $\T$ and two for $\Psi$, so the rank of the kappa-symmetry is eight, namely, $N_\k=N_{\tilde\k}=4$ \cite{Zarembo:2010sg}. The background in this case is $\mathrm{AdS}_3\times \mathrm{S}^3$.
  \item SU$(1,1|1)$ - in this case the eigenvalues of $A$ and $B$ are $\{\a\}$ and $\{\a,0\}$, so there is one solution for $\T$ and one for $\Psi$, so the rank of the kappa-symmetry is four, namely, $N_\k=N_{\tilde\k}=2$. The background in this case is $\mathrm{AdS}_3\times \mathrm{S}^1$.
\end{itemize}
In both cases the kappa symmetry rank is half the number of fermionic d.o.f.

\subsection{Type-To}
The coset background is
\be\label{eq:cosetTo}
\frac{\mathrm{OSP}(M|2N)^2}{\mathrm{SO}(M)\times \mathrm{SP}(2N)}.
\ee
In our classification we are interested only in AdS$_3$ models, so we take $N=1$.
The relevant models in our classification are
$$
\mathrm{AdS}_3 \times \frac{\mathrm{SO}(M)^2}{\mathrm{SO}(M)}\simeq\frac{\mathrm{OSP}(M|2)^2}{\mathrm{SO}(M)\times \mathrm{SP}(2)},\quad
$$

According to \cite{Zarembo:2010sg} we have to find the number of solutions to the equation $A\T=\T B$, so we need to find out how many eigenvalues of $A=\mathrm{SO}(M)$ and $B=\mathrm{SP}(2)$ coincide upon the Virasoro constraint. In general the eigenvalues are different and we don't have kappa-symmetry.
There are three exceptions
\begin{itemize}
  \item OSP$(1|2)$ - in this case the eigenvalues of $A$ and $B$ are $\{0\}$ and $\{0,0\}$, so there are two solutions for $\T$, so the rank of the kappa-symmetry is four, namely, $N_\k=N_{\tilde\k}=2$ \cite{Zarembo:2010sg}. The background in this case is $\mathrm{AdS}_3$. In This case the kappa-symmetry rank is equal to the number of fermionic d.o.f.
  \item OSP$(2|2)$ - that is the same case as SU$(1,1|1)$ of $\mathrm{AdS}_3\times \mathrm{S}^1$ with kappa-symmetry of rank four.
  \item OSP$(3|2)$ - in this case the eigenvalues of $A$ and $B$ are $\{0,\a,-\a\}$ and $\{\a,-\a\}$, so there are two solutions for $\T$, so the rank of the kappa-symmetry is four, namely, $N_\k=N_{\tilde\k}=2$ \cite{Zarembo:2010sg}. The background in this case is $\mathrm{AdS}_3\times \mathrm{S}^3$. In This case the kappa-symmetry rank is one third of the number of fermionic d.o.f.
\end{itemize}

\bibliography{postdoc}
\bibliographystyle{JHEP}
\end{document}